\documentclass[12pt, preprint]{aastex}

\begin{document}

\title{The Chemical Compositions of the Type II Cepheids -- The BL Her and W Vir Variables}

\author{Thomas Maas}
\affil{The W.J. McDonald Observatory, The University of Texas; Austin, TX 78712-1083
\\ thomas$_{-}$maas@hotmail.com}

\author{
Sunetra Giridhar}
\affil{Indian Institute of Astrophysics;
Bangalore,  560034 India\\
giridhar@iiap.res.in}

\author{David L.\ Lambert }
\affil{The W.J. McDonald Observatory; University of
Texas; Austin, TX 78712-1083
\\ dll@astro.as.utexas.edu}

\begin{abstract}

Abundance analyses from high-resolution optical spectra are presented for
19 Type II Cepheids in the Galactic field. The sample includes both
short-period (BL Her) and long-period (W Vir) stars. This is the
first extensive abundance analysis of these variables. 
The C, N, and O abundances with similar spreads for the BL Her and
W Vir show evidence for an atmosphere contaminated with $3\alpha$-process
and CN-cycling products.
 A notable
anomaly of the BL Her stars is an overabundance of Na by a factor
of about five relative to their presumed initial abundances. This
overabundance is not seen in the W Vir stars.
 The abundance
anomalies  running from mild to extreme  in W Vir stars 
but not seen in the BL Her stars
are attributed to dust-gas separation that provides an atmosphere
deficient in elements of high condensation temperature, notably
Al, Ca, Sc,  Ti, and $s$-process elements.
 Such anomalies have previously been seen among RV Tau
stars which represent a long-period extension of the variability enjoyed
by the Type II Cepheids.  
Comments are offered on how the contrasting abundance anomalies of BL Her
and W Vir stars may be explained in terms of the stars' evolution from the
blue horizontal branch.

{\it Subject headings: stars:abundances -- stars:AGB and post-AGB --
stars: variables:other (RV\,Tauri)}

\end{abstract}

\section{Introduction}

Wallerstein (2002)
remarks that  Type II Cepheids
``include most intrinsic variables
with periods between 1 and about 50 days, except for the classical Cepheids and the
shortest semi-regular variables of type M.'' These bounds on the
periods place the Type II Cepheids in the instability strip
 between the RR Lyrae stars at the short
period limit and the
RV Tau variables at the long period limit.
Type II Cepheids fall into two classes:  BL Her stars with periods of 1 to
5 days and the 
 W Vir stars with periods longer than
about 10 days with an indistinct boundary at
about 20-30 days separating these stars from the RV Tau stars.
 Kraft (1972) drew attention
to the period gap (6 to 9 days) between the BL Her and W Vir
stars in globular clusters.

This paper is devoted to abundance determinations of field
Type II Cepheids.
Compositions of these stars
 have received scant attention from spectroscopists despite
hints  of unusual compositions.
Oddly, neither  W Vir nor BL Her, the two prototypes,  have
been subjected to a modern analysis.
Quantitative spectroscopy based on modern model atmospheres and
CCD spectra seems to be limited to
 TX Del (Andrievsky et al. 2002)\footnote{Andrievsky et al. identified
TX Del as a first-overtone Type I Cepheid - see below.}
 ST Pup, a star with RV Tau-like
abundance anomalies (Gonzalez \& Wallerstein 1996),
 and the  C-rich stars V553 Cen and
RT TrA (Wallerstein \& Gonzalez  1996; Wallerstein et al. 2000).
Model atmosphere analyses based on image-tube photographic spectra
were reported for BL Her (Caldwell \& Butler 1978),
$\kappa$ Pav (Luck \& Bond 1989) and
AU Peg (Harris, Olszewski, \& Wallerstein 1984). Curve of growth
analyses and photographic spectra provided abundance estimates
for TW Cap (Anderson \& Kraft 1971), W Vir (Barker et al. 1971), and
$\kappa$ Pav (Rodgers \& Bell 1963, 1968). This very mixed bag of
abundance estimates is an inadequate basis from
which to draw conclusions concerning the evolution of Type II Cepheids
and other potential origins for abundance anomalies.

The principal prior indication of an anomalous composition for Type II Cepheids
concerns 
the reported
 deficiency of the $s$-process elements (relative to the iron abundance) 
--
see Rodgers \& Bell (1963, 1968),  Barker et al. (1971) and
especially  Luck \& Bond (1989). Our interest was driven in part by this
indication, but also by the lack of a thorough study of field Type II Cepheids, and
finally by their close relationship to the RV Tau stars for which abundance
anomalies exist.
The   principal 
anomaly for some
 RV Tau stars is one in which the atmosphere is deficient in those
elements that first  condense into grains as gas is cooled, i.e.,
there has been a separation or winnowing of dust from gas and
accretion of gas by the star (Giridhar et al. 2005). An apparently
rarer anomaly is one in which elements with a low first ionization
potential are underabundant (Rao \& Reddy 2005). Other RV Tau
stars have a normal composition.
There is ample evidence that the W Vir and RV Tau stars have much in common.
For example, photometry of Type II Cepheids and RV Tau stars
 in the LMC  shows clearly
that the RV Tau stars and the Type II Cepheids define a common
Period-Luminosity-Color relation (Alcock et al. 1998).
Given this commonality, the question naturally arises --
What abundance anomalies are shown by the BL Her and W Vir stars
and are those anomalies similar to those shown by the RV Tau stars?

The observations and abundance analyses of 19 Type II Cepheids
 are described in Section 2.
Results for the individual stars and remarks on their classification
as Type II Cepheids are provided in Section 3. Discussion of the possible
anomalous abundances and their relation to the prior evolution of the
 stars and to other processes is provided in Section 4. Brief remarks on  
the spectroscopic calibration of photometric  measurements of
metallicity are given in Section 5. Concluding remarks
are offered in Section 6.

\section{Observations and Analyses}

The program stars were observed with
 the W.J.  McDonald Observatory's 2.7m Harlan J. Smith reflector and the
CCD-equipped `2dcoud\'{e}' spectrograph (Tull et al. 1995) in observing runs 
in 2004 and 2005. A spectral
resolving power
$R = \lambda/\Delta\lambda \simeq  60,000$ was used and a broad spectral range was covered
in a single exposure.

Spectra were rejected
if they showed line doubling, markedly asymmetric lines, or
strong emission in the Balmer lines. 
 It is presumed that the spectra not
showing these characteristics represent the atmosphere at a
time when standard theoretical models may be applicable.
This presumption should be tested by analysis of a series of spectra taken
over the pulsational cycle. This remains to be done but 
we have analysed three stars using spectra taken at
different phases and obtained consistent results.
The program stars and dates of observation are listed in Table 1.

The abundance analyses were performed as described in our
earlier papers on the RV Tau variables. Atmospheric parameters are
determined  from the Fe\,{\sc i} and Fe\,{\sc ii}
lines by demanding excitation and ionization equilibrium, and
that the iron abundance be independent of the equivalent width.
The adopted parameters listed in Table 1  were determined using
Kurucz model atmospheres and a normal helium abundance (He/H = 0.1).
The full abundance analysis used a Kurucz atmosphere
with the parameters in Table 1.
 Abundances are
referred to the recommended solar photospheric abundances
given by  Asplund, Grevesse, \& Sauval (2005).
Table 1 shows the anticipated average difference in atmospheric
parameters
 ($T_{\rm eff}, \log$ g)
 between the BL Her
and W Vir stars: 
 ($T_{\rm eff}, \log$ g) $=$ (5970,1.4) and (5380, 0.6) for the BL Her and
W Vir stars, respectively.

\section{The Program Stars}

Harris (1985) provided what is regarded as the reference catalog of
Type II Cepheids. His principal criterion for distinguishing a Type II
from a Classical (Type I)
  Cepheid was distance from the Galactic plane ($|$Z$|$). Our selection
of stars was based on
his  catalog but includes some recent discoveries. Various
photometric indices extracted from the light curves have been
proposed as distinctive marks of a Type II or a Type I Cepheid but
the distinction remains tricky, as many authors remark. In addition,
longer period 
Type II Cepheids overlap in pulsational properties (i.e., period and
luminosity) with the RV Tau variables. Therefore, not only must
population type be examined but variable type too. In the following
text, we comment on the our selection of Type II
Cepheids and remark on those with RV Tau proclivities.

In this paper, the distinction between a BL Her and a
 W Vir variable is maintained; the reasons for this will become
obvious when the compositions are presented. Here, a BL Her
variable is one with a pulsation period of four days and shorter,
and a W Vir variable is a star with a period of 11 days or longer.
Four stars in the sample are of intermediate period with
two judged by composition to be BL Her stars and two W Vir
stars. 

The stars are discussed  in order of increasing pulsation period.

{\bf BX Del:} BX Del with an estimated $|$Z$|$ of 0.3 kpc
was listed as a Type II Cepheid by Harris (1985). 
The
metallicity [Fe/H] $= -0.2$ places BX Del at about the upper limit for
thick disk stars (Reddy et al. 2006).

{\bf VY Pyx:} This variable was identified as a Type II Cepheid
by Sanwal \& Sarma (1991).
The composition shows similarities with that of BX Del.

{\bf BL Her:} This Type II Cepheid is the prototype of the subclass of
the shorter period  variables.

{\bf SW Tau:} The light curve of this bright BL Her star is that of a
Type II Cepheid. The metallicity, [Fe/H] of $+0.2$,  places SW Tau
as the most metal-rich star in our sample.
 SW Tau is C-rich with the
C\,{\sc i} lines  strong by inspection (Figure 1).
This figure shows a portion of  the
spectrum of SW Tau, IX Cas (a very C-poor star),
 CC Lyr 
(a star of normal C abundance but having an extreme Fe-deficiency),
and BL Her. The four stars have similar atmospheric parameters and, therefore,
to first order the variations in the strength of a given line from
star-to-star reflect abundance differences between the stars.

{\bf AU Peg:} AU Peg is a single-lined spectroscopic binary (Harris,
Olszewski, \& Wallerstein 1984) with an orbital period of
53.3 days. Harris et al. argue that the secondary is a more
massive compact object.
A feature of AU Peg is
that it lies close to or just beyond the red edge of
 the instability strip (Harris et al 1984; Vink\'{o}. Szabados, \&
Szatm\'{a}ry 1993).
 Presently, AU Peg is close to filling its
Roche lobe and mass transfer between the two stars almost certainly
occured at an earlier time. Dust, as indicated by an IR-excess (McAlary \&
Welch 1986), surrounds the binary. Outflow from the system is suggested
by the appearance of  P Cygni-like feature accompanying AU Peg's
H$\alpha$ profile (Vink\'{o} et al. 1998).

{\bf DQ And:} Harris (1985) listed this star as a Type II Cepheid
on the basis of the estimated distance from the Galactic plane:
$|$Z$| = 0.6$ kpc. 
 Balog, Vink\'{o}, \& Kasz\'{a}s (1997),
put
$|$Z$|$ at 2.3 kpc. The radial velocity of $-231$ km s$^{-1}$
(Harris \& Wallerstein 1984) is certainly not that expected of a Pop. I
star.
However, the Baade-Wesselink radius  
puts DQ And on the period-radius (P-R) relation defined
by Classical Cepheids (Balog et al. 1997).
The star is here considered to be of Type II.

{\bf UY Eri:} With a [Fe/H] of $-1.8$ and
a radial velocity of 171 km s$^{-1}$ (Harris \& Wallerstein 1984),
 UY Eri is obviously a Type II Cepheid
and member of the Galactic halo.
Its composition is typical of a halo star with [Fe/H] of $-1.8$
(McWilliam 1997).

{\bf TX Del:} Harris (1985) includes the star in his table of
Type II Cepheids. Harris \& Welch (1989) found  that the star
is a single-lined spectroscopic binary with an orbital period of
133 days. Schmidt et al. (2005) show that the star has varied in
mean brightness, a characteristic not associated with a Classical
Cepheid. Andrievsky et al. (2002) consider TX Del to be a
first-overtone Classical Cepheid.\footnote{Our derived abundances
are in quite good agreement with those by Andrievsky et al. (2002);
the mean difference in absolute abundance from 12 elements in common is
$-0.15$ dex.}
 Although this identification may
ease an explanation for the star's  solar metallicity,
it places this Pop. I object more than  1 kpc from the
Galactic plane. The Baade-Wesselink radius (Balog et al. 1997)
places TX Del, as it does DQ And,
 on the P-R relation for Classical Cepheids. 
An explanation as a runaway star  seems
unlikely given that TX Del is a binary. We consider TX Del
to be a Type II Cepheid and recall, as did Harris \& Welch, the
association of Pop. II characteristics and near-solar metallicity
held by metal-rich RR Lyrae stars.

{\bf IX Cas:}  A Type II star according to Harris (1985) who gave
$|$Z$|$ as 0.7 kpc. This is the primary of a spectroscopic binary (Harris \&
Welch 1989).
The Baade-Wesselink radius puts the star on the Type II P-R
line. 
The pulsation period of 9.2 days is long for a BL Her star but the
high Na abundance which  distinguishes BL Her from W Vir stars
places this star among the BL Her class.
The  outstanding mark of IX Cas's spectrum  is
the weakness of the C\,{\sc i} lines.
 Figure 1 shows that the C\,{\sc i} lines
typically used in our analyses are weak or absent from the spectrum of
IX Cas. Lines near 9050 \AA\ which are normally too strong for use in the
abundance analysis
had to be used in the case of IX Cas.
The low C abundance is
reminiscent of values found for the
weak G-band giants (Sneden et al. 1978).

{\bf AL Vir:} AL Vir is a Type II Cepheid according to distance from the
Galactic plane (Harris 1985), light curve (Schmidt et al. 2004a), He\,{\sc i}
5876 \AA\ emission (Schmidt et al. 2004b), and the Baade-Wesselink
radius (Balog et al. 1997).
AL Vir with a pulsation period of 10.3 days marks the transition from
BL Her to W Vir stars in our listing of variable by increasing period.
The composition (particularly the lack of a Na
overabundance) suggests AL Vir is a W Vir star.

{\bf AP Her:} Harris (1985) considered AP Her a Type II Cepheid on the
basis of the estimated $|$Z$|$ of 0.4 kpc. 
At [Fe/H] $\simeq -0.8$, this may be considered a thick disk star.
The literature on
Type II Cepheids is peppered with remarks about the difficulties in
distinguishing between Type I and II Cepheids. This is well illustrated
by Schmidt et al. (2004a) who, in a discussion of stars with large period
changes, put  AP Her with stars that `are likely type II Cepheids'
but later place 
AP Her  among the variables
 `for which the predominance
of the evidence indicates  type I classification'.
The [Fe/H] 
confirms that AP Her is not a Classical Cepheid.

{\bf CO Pup:} The literature on this star is extremely sparse. Following
Harris's (1985) listing of CO Pup as a Type II Cepheid, the star has
featured in just five papers (according to SIMBAD), all providing
or regurgitating photometry with not a single spectroscopic observation
reported. There appears to be confusion as to whether it is a Type I or
II star. Our estimate of the intrinsic [Fe/H] of $-0.6$ 
suggests that it is not a Classical
Cepheid.

{\bf SZ Mon:} This Cepheid is not listed by Harris (1985). Apparently, it
was assumed to be a Classical Cepheid. This assumption
was rejected
 by Stobie (1970) and Lloyd Evans (1970), largely on the
grounds that alternating minima are of different depths. The star is either a
W Vir or a RV Tau variable. If the latter description
is correct, the fundamental period is approximately 32.6 days.
SZ Mon has a marked infrared excess  
 (McAlary \& Welch 1986).

{\bf W Vir:} The prototypical Type II Cepheid with an extensive record of
published photometric and spectroscopic observations.
As befits the prototype, its composition is broadly representative of the other
W Vir stars. Our results are in fair accord with the first results
given long ago by Barker et al. (1971).

{\bf MZ Cyg:} Type II according to Harris (1985) with $|$Z$|$ of 0.9 kpc.
The light curve (Schmidt et al. 2004a) is unlike that of a Classical
21 day Cepheid.  Emission in H$\alpha$ and He\,{\sc i} 5876 \AA\ occuring
coincident with the bump on the rising light curve 
is taken to 
be a discriminant between Type II and Type I Cepheids. Schmidt
et al. (2004b) report such emission for MZ Cyg and thus
confirm its designation as a Type II Cepheid. 
The [Fe/H]  reminds one of the metal-rich RR Lyraes with
Pop. II kinematics.

{\bf CC Lyr:} Harris (1985) estimated a $|$Z$|$ of 2.7 kpc
 and, hence, identified it as a Type II Cepheid.
Photometry shows alternating
minima of different depths, a characteristic of RV Tau variables
(Schmidt et al. 2004a). CC Lyr has a strong infrared excess, also
invariably a characteristic of RV Tau variables. 
This 
star's spectrum betrays an obvious signature
 of dust-gas separation, a mark of warm RV Tau variables.
CC Lyr's spectrum
contains  few lines, a reflection of the fact that the atmosphere
is highly depleted in many elements including Fe.
Figure 1 shows C\,{\sc i} lines but not  the Fe\,{\sc i}
and Ni\,{\sc i} lines seen in spectra of the other three stars.
The intrinsic metallicity is about [A/H] $\simeq -0.8$ as indicated by the S and Zn
abundances which we assume are undepleted. The measured iron abundance is
[Fe/H] $\simeq -4$ which corresponds to a dust-gas depletion of more
than three orders of
magnitude!

{\bf RX Lib:} Harris (1985) put RX Lib in his table; the estimated
$|$Z$|$ was 3.2 kpc. Earlier, Harris \& Wallerstein (1984) placed RX Lib in
a category `Stars thought not to be Cepheids' remarking that the
type for this star is `uncertain'. Apparently, observations -- spectroscopic
and photometric -- have not been published since the mid-1980s on 
this star. 
Although the type may be `uncertain', RX Lib's composition is 
similar to that of W Vir.

{\bf TW Cap:} The metallicity, [Fe/H] $= -1.8$, and essentially
a normal composition for a halo star suffice to identify this as
a Type II Cepheid. The star's position in the P-R diagram
supports the Type II designation (B\"{o}hm-Vitense et al. 1974).
 Harris (1985) classified the star thus on the
basis of its estimated 2 kpc distance from the Galactic plane.

{\bf V1711 Sgr:} Harris (1985) listed this star as a Population II
Cepheid. In the limited literature on V1711 Sgr, there is
a lone dissenting voice about the classification. Berdnikov \& Szabados (1998)
find a difference in amplitude between their light curve and that obtained
about two decades earlier by Dean et al. (1977) and suggest that the
star may be a SRd variable. 
Presence of an IR excess (Lloyd Evans 1985; McAlary \& Welch 1986;
Smith 1998) and a composition bearing
the signature of dust-gas separation (see below) together
with the long period suggest that the star may be related to the
RV Tau variables.
We follow Harris and include the star among our sample of Type II
 Cepheids. The period of 28.56 days reported Berdnikov \& Szabados (1998)  
is adopted and not the 30.5 days listed by Harris (1985).

\section{Implications of the  Chemical Compositions}

\subsection{The Evolutionary Context}

Type II Cepheids in the instability strip above the horizontal
branch have evolved from stars on the 
blue  end of the horizontal branch (BHB) where the stars are He-core burners. 
Placement of a star on  the horizontal branch
following evolution to the tip of the first red giant branch
requires substantial mass loss by processes
as yet unidentified.
Seminal calculations about Type II Cepheids
 were reported by Gingold (1974, 1976, 1977, 1985).
Gingold (1985, Figure 1) shows two evolutionary tracks starting from
the BHB that may bracket the
origins of the Type II Cepheids. Figure 2 is an adaptation of Gingold's
figure.
The form of the track appears to depend principally on the
mass of the envelope (relative to the core mass) and weakly on the
initial metallicity.

In the simpler of the two tracks,   the blue horizontal branch
star evolves to the red, crosses  the instability strip as a BL Her variable,
and evolves up the AGB to endure thermal pulses
 before leaving the AGB
tip for rapid evolution to the blue across the instability
strip as a RV Tau star and  a brief experience as
as a post-AGB star. We refer to this type of evolution as following a track-direct.

In the alternate path, a star from the  more extreme blue part of
the BHB
evolves to the red, crosses the instability strip to
approach the AGB but  experiences a structural readjustment
between the H and He shell burning shells that directs the evolutionary
track back to the blue. The star executes a `bluenose' (Gingold's
parlance) involving two more transits across the instability strip
and returns to the AGB from which after increasing in luminosity
 it can make
its final departure across the instability strip to the blue as a
post-AGB star. This track makes four crossings of the instability
strip: the first three with a period representative of BL Her variables and 
the final one as a W Vir or RV Tau variable. 
We refer to this type of evolution as following a track-bluenose.

In both  evolutionary tracks, the
luminosity difference between  the earlier one or three  and the final
crossings of the instability strip is the key to  
the period gap in the distribution function of
Type II Cepheids between the BL Her and W Vir
variables. This distribution necessarily  depends
on the  time taken to transit the instability strip on each
crossing
(Gingold 1985) and on the rate of production of stars evolving along a
track-direct and a track-bluenose.

Metal-rich RR Lyraes discovered by Preston (1959) long posed
a puzzle because
metal-rich low-mass stars after the He-core flash
settle as He-core burning giants to the red of the RR Lyrae gap, the
so-called clump giants.
A clump He-core burning
 giant evolves   up the AGB where
episodic thermal pulses begin at high luminosity.
 Such pulses
are not predicted
to lead to excursions into the instability strip to the
blue of the AGB (see Iben 1991 - his Fig. 5 and Gingold 1985 - his Fig. 1).
 Taam,
Kraft, \& Suntzeff (1976) proposed that severe mass loss would
place  giants with a low mass envelope away from the clump and on the 
horizontal branch in or to the blue of the RR Lyrae gap.
Subsequent
evolution of these horizontal branch stars 
produces a metal-rich BL Her or W Vir, as above.
  Gingold (1977)
also showed that even stars placed on the red horizontal
branch would, if the envelope mass were reduced sufficiently, evolve 
first to the blue. 
Taam et al.  showed that kinematically the metal-rich RR Lyraes were
members of the old (now, thick) disk. 

How thick-disk giants lose the required amount of mass (about $0.5M_{\odot}$
-- Taam et al. 1976) and reduce their envelope to a few per cent of a solar
mass is  unknown. Observers may speculate that the He-core flash that
precedes He-core burning may lead in rare but sufficient cases to
internal violence and mass loss. This oft-invoked speculation
has yet to find a resonance in theoretical calculations -- see recent
reports by Deupree (1996) and Dearborn, Lattanzio, \& Eggleton (2006).
In the absence of observationally confirmed 
theoretical ideas on how severe mass loss
is achieved, it is impossible to predict the changes of surface
composition expected of BL Her, W Vir, and RR Lyr stars relative
to their initial (main sequence) composition.
Horizontal branch stars including RR Lyr variables must be  affected
by the first dredge-up occuring at the base of the
red-giant branch and by other subsequent events such as the
mass loss required to account for a star's position on the
horizontal branch. Principal effects will be on C, N, and O,
possibly He, and certainly Li, Be, and B.

Three BL Her stars in our sample (IX Cas, TX Del, and AU Peg)
and the W Vir star ST Pup (Gonzalez \& Wallerstein 1996) are known spectroscopic
binaries for which previous mass transfer may have directed their
primary star to the instability strip. Although other binaries
 may lurk undetected
in the sample, well observed stars like W Vir must, if accompanied
by a companion, have a very low velocity amplitude.

\subsection{Observed composition and period}

The BL Her stars and the
W Vir stars occupy different positions
 along the evolutionary tracks from the blue horizontal
branch to the post-AGB departure from the AGB;  the BL Her stars are
closest to the blue horizontal branch and the W Vir stars are presumed
to be leaving the
AGB. In light of this difference in staging along the evolutionary
tracks,
we distinguish the two classes in discussing their
compositions. 

The BL Her variables are stars with pulsation periods of four
days or shorter plus TX Del and IX Cas, stars of intermediate
period but with the  signature of BL Her stars, i.e, an overabundance of
Na. The BL Her stars have on average a higher [Fe/H] than the W Vir
stars.
The W Vir variables are stars
with pulsation periods of greater than about
 10 days plus the intermediate
period stars AL Vir and AP Her.
Among the W Vir stars  is CC Lyr which, as noted
above, has a composition indicative of severe dust-gas separation,
a common feature of RV Tau variables. Other W Vir stars offer
indications of dust-gas separation and some
have light curves suggestive of RV Tau-like pulsations. 

The compositions of the stars are given in Table 2 for the
BL Her stars, Table 3 for the intermediate period quartet, and
Table 4 for the W Vir stars. In Figures 2 to 5 displaying the results
as [X/Fe] versus [Fe/H], the BL Her stars (including TX Del and IX Cas)
are represented by unfilled circles, the W Vir stars (including
AL Vir and AP Her) by filled circles. In addition to our sample, we plot
the previously published results for
the C-rich BL Her stars  -- V553 Cen (Wallerstein \& Gonzalez 1996)
and RT TrA (Wallerstein, Matt, \& Gonzalez 2000) -- as unfilled
squares, the W Vir stars ST Pup (Gonzalez \& Wallerstein 1996) 
and $\kappa$ Pav (Luck \& Bond 1989) as filled triangles.

Studies of the kinematics of Type II Cepheids show that the examples
in the solar vicinity are primarily  thick disk stars with
an admixture of halo stars. 
Initial compositions of thick disk and halo stars are now known
as a function of metallicity.
The thin and thick disk do not have identical compositions
as a function of metallicity (Bensby et al. 2005; Reddy et al.
2006). The differences  at a given [Fe/H] are not large. 
Of relevance to our discussion is the fact that at a given [Fe/H]
the scatter in initial abundance ratios, i.e., [X/Fe], is very small
for both the thin and the thick disk, at least for the stars now in the
solar neighborhood (Edvardsson et al. 1993; 
Reddy et al. 2003, 2006). At a common  low [Fe/H], the composition of
thick disk and halo stars appear to merge but the scatter in [X/Fe]
at a given [Fe/H] may be more significant for the halo population. Since
our sample includes just two halo stars, this scatter
in composition, if real, is unimportant here.

The compositions of the variables are judged relative to their
presumed initial compositions as inferred (usually) from their
[Fe/H], but, an alternative is considered in the cases where
dust-gas separation is suspected.

\subsection{The BL Herculis variables}

Our sample is augmented by published analyses of the 
BL Her C-rich variables V553 Cen and RT TrA.\footnote{Other 
 C-rich variables
listed by Lloyd Evans (1983) have periods of 20 days or longer. None
have been subject to quantitative analysis.}
Consideration of the C, N, and O abundances led Wallerstein
and colleagues to propose that the atmospheres of the
two C-rich variables had been enriched
in $^{12}$C from the $3\alpha$--process followed
by operation of the H-burning CN-cycle.
The high C abundances pointed to the $3\alpha$-process. Operation of the
CN-cycle was indicated by the high N abundances.
Measurement of the $^{12}$C/$^{13}$C ratio for both stars
gave results
 equal to the equilibrium value for the CN-cycle, an indication that
CN-cycling occurred following the $^{12}$C enrichment from the
$3\alpha$-process.
A Na enrichment was attributed to proton capture on $^{22}$Ne 
occuring at the time of  H-burning.
This combined operation of the H-burning CN cycle and the He-burning $3\alpha$-process
and attendant  reactions may account for the BL Her stars
analyzed here.  (The halo variable UY Eri is not discussed
further.)

Prior to a star's transfer to the horizontal branch, the first dredge-up
increased the surface N abundance at the expense of the C
abundance. The maximum possible N abundance post dredge-up is obviously
equal to the sum of the initial C and N abundances. Oxygen is not
predicted to be reduced by the dredge-up. The C and N
abundances show clearly that the first dredge-up has not been the
primary influence on the stars' surface abundances. The C
abundance for all but one star (IX Cas) equals or exceeds the initial
abundance. The N abundance except for IX Cas and DQ And exceeds by
about 0.6 dex that
predicted by complete conversion of initial C and N to N.
The N abundance is approximately that
expected by complete conversion of initial C, N, and O to N but the
O abundance is not depleted but has an abundance similar to or
even slightly greater than the presumed initial abundance.
One star -- SW Tau -- is C-rich with a C/O ratio of about 3.  A
second -- BL Her -- is borderline C-rich with C/O of 0.9. With
the remarkable exception of IX Cas (C/O $= 0.005$), SW Tau, and BL Her,
 the other BL Her
stars have a C/O ratio in the range 0.1 to 0.4.
The C, N, and O abundances  signal the presence of
$3\alpha$-process  and CN-cycle products.

A sodium overabundance 
is an obvious feature of the eight BL Her variables 
 (Figure 3).
The [Na/Fe] ratio over the interval $-0.5 <$ [Fe/H] $< +0.2$
is independent of [Fe/H] with a spread of about 0.5 dex and
 a mean [Na/Fe] $= +0.73$, a
substantial increase over the ratio of $+0.12$ for the
thick disk (Reddy et al. 2006). Wallerstein and colleagues
obtained a  Na overabundance for their pair of
C-rich BL Her variables: [Na/Fe] =  $+0.74$ (RT TrA) and $+0.43$ (V553 Cen).
The mean [Na/Fe] for the W Vir stars excluding the halo star TW Cap, and
the stars (ST Pup, CC Lyr, and V1711 Sgr) affected by dust-gas separation is
[Na/Fe] $= +0.23$, a value consistent with the ratio of unevolved thick
disk stars.

The Al abundances of the BL Her stars
give a mean 
[Al/Fe] $= +0.13$, a value confirmed by the pair of C-rich
variables. The mean [Al/Fe] is 
slightly less than that of thick disk stars for which Reddy et al. (2006)
gave [Al/Fe] $= +0.30$.  The mean [Na/Al] of the BL Her stars is
$+0.60$ but the presumed initial ratio for the thick disk stars
is $-0.18$, a factor six change in the Na/Al ratio occuring in the
course of evolution to a BL Her variable. The W Vir stars show a
wide range in Al abundance  which we attribute to 
dust-gas separation but the upper limit in the range is consistent
with the ratio for thick disk stars.

The $\alpha$-elements, here Mg, Si, S, Ca, and Ti, conform to
expectation. At thick disk metallicities,
 the ratios [$\alpha$/Fe] are positive for the dwarfs:
[$\alpha$/Fe] =  $+0.32$ (Mg), $+0.22$ (Si), $+0.18$ (Ca), and
$0.21$ (Ti)
 at [Fe/H] $\simeq -0.5$ (Reddy et al. 2006).
 (Sulphur was not examined by Reddy et al. but a value
for [S/Fe] close to those of Mg and Si is expected.) Thick disk stars
with [Fe/H] $> -0.3$ have thin disk abundances with a decline of [$\alpha$/Fe]
to zero at about [Fe/H] = 0. 
The observed [$\alpha$/Fe] ratios versus [Fe/H] are shown in Figure 4. They
 match expectation well for the composite
index from Mg, Si, and S. The observed [Ca/Fe] for [Fe/H] $< 0$
stars is as expected but for the low [Ca/Fe] for AU Peg.
AU Peg also shows a low [Ti/Fe].

Iron-group abundances (here, Cr, Mn, and Ni) shown in Figure 5
as a function of [Fe/H] follow well the trends expected of
disk stars (Reddy et al. 2006):  [Cr/Fe] $\simeq$ [Ni/Fe] $\simeq 0$
and [Mn/Fe] $\leq 0$.
Two of the stars -- BX Del and TX Del -- appear to be Sc-poor
relative to expectation. Both are also Ti-poor.
 The Zn abundances follow expectation,
i.e., [Zn/Fe] $\simeq 0$, with the possible exception of
DQ And.

Abundances of heavy elements in Type II Cepheids  have occasioned
considerable comment beginning with Rodgers \& Bell's (1963)
analysis of $\kappa$ Pav where they reported an underabundance
by about a factor of seven of $s$-process elements (relative to
the Fe abundance). A new analysis of $\kappa $ Pav was made by
Luck \& Bond (1989) who found [$s$/Fe] $= -0.4$ at [Fe/H] $= 0.0$.
The $s$-process contribution is here compiled as a simple mean of the
derived Y, Zr, La, and Ce abundances. Neodymium with roughly equal
$s$- and $r$-process contributions at the solar composition (Burris et al. 2000)
was not included but its inclusion would not alter any
conclusions. The expected [$s$/Fe] for the thick disk is
close to zero. Figure 6 shows that this expectation is
found for five of our seven BL Her variables, and 
C-rich V553 Cen, and RT TrA. Three BL Her stars -- BX Del, VY Pyx,
and TX Del -- have [$s$/Fe] $\simeq -0.5$, a value far from
the range shown by thick (and thin) disk unevolved stars. 
There is a tantalising hint that the [$s$/Fe] values offer a
bimodal distribution. The low-$s$ stars are discussed
further in the next section on the W Vir stars for which
underabundance of $s$-process elements is common.

Europium is taken as the sole measure of the $r$-process
contribution to the stellar composition; Eu in the solar
composition is about 97\% $r$-process (Burris et al. 2000).\footnote{Luck
 \& Bond
adopted the mean of their measured elements from La to Eu
 as an indicator  of the
$r$-process. This
choice includes La and Ce which in the solar mixtures
are
about 80\% $s$-process and  only 20\% $r$-process.}
Europium abundances for the BL Her  variables
are shown in Figure 7. A tight relation is seen with a
tendency for [Eu/Fe] to increase  with decreasing [Fe/H].
This relation mimics that for the thick disk but is displaced
at a given [Fe/H] to lower [Eu/Fe] by about 0.2 dex. 
One presumes that this is an artifact of the analysis and not a
consequence of stellar evolution.

Among the W Vir stars,  dust-gas separation is
unmistakably severe in the case of CC Lyr and suspected in
several other cases. The signatures of severe dust-gas separation
are [S/Fe] and [Zn/Fe] ratios in excess of their normal
ratios for thick disk stars
 of about $+0.3$ and 0.0, respectively, and negative 
ratios for [Al/Fe], [Sc/Fe], and [Ti/Fe]. 
In mild cases,  the effects of separation are limited to the
elements of highest condensation temperature (Al, Sc, and Ti) and, then 
[Al/Fe], [Sc/Fe], and [Ti/Fe] show negative values but [S/Fe] and [Zn/Fe]
have  normal values. 
In our sample of BL Her variables, none show the signature of
severe dust-gas separation. The trio with a low [$s$/Fe], also a
signature of depletion of the highest condensation
temperature elements, provide weak hints of depletion of Al, Sc, and Ti.

In summary, the composition of the BL Her variables, as suggested by Wallerstein
and colleagues from their analyses of the C-rich pair of variables, has
been set by mixing with the products of $3\alpha$- processing, 
CN-cycling, and $p$-capture on $^{22}$Ne. Other abundance ratios relative to
Fe have their expected values with the  exception of low $s$-process
abundances for a minority of the group.

\subsection{The W Virginis Variables}

The W Vir stars as a class show evidence in their C, N, and O
abundances for the presence of $3\alpha$-process and CN-cycle
products in their atmosphere but not the Na overabundance
seen for the BL Her variables. A signature of weak dust-gas
separation seems present in some stars with
a clear signature of severe separation present for
CC Lyr and ST Pup (Gonzalez \& Wallerstein 1996).
The sample of W Vir stars has on average a lower (intrinsic) [Fe/H]
than the  sample of BL Her stars.

The first dredge-up is clearly not the defining part of the evolutionary
history. For four stars, although the N abundance may be equal to
the sum of the initial C and N abundances, the C abundance either
remains equal to the initial abundance or exceeds it by as much as
0.8 dex. In two cases, the N abundance exceeds the sum of the
initial C and N abundances. The C/O ratios range from
about 2 for MZ Cyg and SZ Mon to 0.06 for W Vir.
 Although lacking a
counterpart to IX Cas with its low C abundance, the spread in
C, N, and O abundances is similar to that of the BL Her variables.

The striking difference in the [Na/Fe] ratios of the BL Her and
W Vir stars  (Figure 3) is a clue to their different origins.
Setting ST Pup and V1711 Sgr
aside, the [Na/Fe] of the
W Vir stars appears independent of [Fe/H] and pulsation period, and consistent
with that found for the thick disk dwarfs but significantly
less than that of the BL Her variables.
The exceptions, both with the high [Na/Fe] representative of the BL Her
variables, are ST Pup and V1711 Sgr. ST Pup (Gonzalez \& Wallerstein 1996) is
seriously affected by dust-gas separation such that neither the present Fe and
nor most likely also the Na abundance are unaffected. It is impossible at
present to correct the observed [Na/Fe] for the effects of dust-gas
separation but a downward revision is certain.
V1711 Sgr may be a Na-rich W Vir star but distortion of the [Na/Fe]
ratio by dust-gas separation effects cannot be excluded.
Although V1711 Sgr is not as dramatically affected by dust-gas
separation as ST Pup, the [Na/Fe] may have been increased
sufficiently (say, by 0.3 dex) to place it among the BL Her stars.
 
Aluminum is in the main less abundant in W Vir than in BL
Her stars (Figure 3). All W Vir stars show a [Al/Fe] less than
that of the thick disk dwarfs. ST Pup's positive [Al/Fe]
is anomalous for a star strongly affected by dust-gas separation
(see below). 

Abundances of the $\alpha$-elements (Figure 4) show a
greater spread for [Ca/Fe] and [Ti/Fe] versus [Fe/H] than is the case for the
BL Her stars. The spread is greater for [Ti/Fe]
than for [Ca/Fe] and almost absent for [Mg/Fe], [Si/Fe], and
[S/Fe]. The appearance of negative [Ca/Fe] and [Ti/Fe] is consistent
with the operation of dust-gas separation. 

The $r$-process element Eu defines a similar  [Eu/Fe] versus [Fe/H]
relation for the BL Her and W Vir stars  which incorporates the two C-rich
BL Her stars (Figure 6).

The $s$-process abundances show apparently bimodal values of
[$s$/Fe] of about 0.0 or $-0.7$, a tendency similar to that
provided by the BL Her stars. Given that [$s$/Fe] $\simeq 0.0$
is representative of the thick disk, the anomalous stars
are those with negative [$s$/Fe]  values. 
As noted above, the $s$-poor stars have negative [Ti/Fe] and not the
positive values expected of thick disk stars and negative [Sc/Fe] with
values of about $-1$ in four cases. (The W Vir star $\kappa$ Pav, as
analysed by Luck \& Bond (1989),  is a typical $s$-process deficient
member of the class.)

Longer-period W Vir stars and shorter-period  RV Tau stars  overlap in
period and, therefore, it of interest to comment on similarities and
differences in composition. 
The C, N, and O abundances of the two groups have very similar
ranges and mean values (Giridhar et al. 2005).
Of especial interest in the [Na/Fe] ratio of
RV Tau stars. Since many RV Tau stars show a Fe-depletion from dust-gas
separation, we compare not [Na/Fe] but [Na/Zn] because Zn is generally not
depleted and [Zn/Fe] $\simeq 0$ in unevolved stars. 
 We compile [Na/Zn] for RV Tau stars
 from results given by Giridhar et al. (2005)
and earlier papers in that series. The mean value from 21 stars is
[Na/Zn] $= +0.17\pm0.03$ to be compared with [Na/Zn] $= +0.20$ for our
W Vir stars and $+0.85$ for our BL Her stars. Excluded from our sample of
RV Tau stars are two stars with the high [Na/Zn] characteristic of the BL Her
stars, CE Vir where the abundance anomalies are not due to dust-gas
separation (Rao \& Reddy 2005), and EP Lyr where the dust-gas separation is
so severe that Na is somewhat depleted. The conclusions are that the
distribution of [Na/Zn] ratios  (i.e., the [Na/Fe] in the absence of dust-gas separation) of the W Vir
 and the  RV Tau stars 
 are identical with the majority showing no sodium enrichment but a
minority of perhaps 10 per cent showing the sodium enrichment of the
BL Her stars.

\subsection{Accounting for the anomalies}

Introduction to  the stellar surface of products of internal
nucleosynthesis is a well known phenomenon accounting in principle  
for a surface composition differing from the 
presumed initial composition. The classic case must be that
of those S stars with Tc present at the surface. The
qualifying phrase `in principle' recognises that the details
of the internal nucleosynthesis and dredge up to the
surface may not yet be understood. Often, the particular processes of
nucleosynthesis involved are identifiable and it is the modes of transport of
the nucleosynthetic products to the atmosphere that are in
doubt.

Such would appear to be the case for the BL Her stars. The
 atmospheres are contaminated,
 as Wallerstein and colleagues first showed from their
analyses of two C-rich BL Her stars, with
$3\alpha$-processed material followed by  exposure to
the CN-cycle.
This recipe which accounts in principle for the C, N, O, and
Na abundances of the BL Her stars
has yet to be
incorporated into a stellar evolution model that ties
the nucleosynthesis and mixing to  events in the life of
a low-mass star. 
The sole remaining abundance anomaly is the appearance
of a low [$s$/Fe] in three stars. There are hints of anomalously
low Ca, Ti, and Sc in a minority. Such hints lack an obvious
nucleosynthetic explanation, even in principle.

In searching for the origin of the abundance anomalies of the
Population II Cepheids, one may begin with the Na  abundance
difference between the W Vir and  BL Her stars. 
Recall that the W Vir stars are considered to have evolved  from
BL Her stars. Sodium destruction is highly unlikely to have
occured along the track between the BL Her and W Vir phases. 
Thus, it seems clear that the W Vir stars observed here
cannot be direct descendants of the BL Her stars observed here;
V1711 Sgr may be an exception.
Since, sodium production is highly unlikely to be a feature of
the short evolutionary phase from the BHB to the instability strip, the
overabundance of sodium in BL Her stars
 must be attributed to the as yet
unknown processes involving mixing and mass loss
 that placed the BL Her progenitors on the BHB.
Continuing this thought we conjecture that the processes
which placed the W Vir progenitors on the BHB did not
provide for a sodium overabundance. 
The unknown processes in both cases involved $3\alpha$-process and
the CNO-cycles to similar degrees. Given that the distributions of
 C, N, and O abundances are
similar for BL Her and W Vir stars, the difference involving the unknown
processes would appear to have
been the temperature at which the CNO-cycles operated.  Sodium 
production by proton capture on $^{22}$Ne occurs at `high' but not
at `low' temperatures. We speculate that mixing and mass loss involving
CNO-cycling at high temperatures placed the star near the red end on the
BHB with the star evolving along a track-direct to become a Na-rich
BL Her star. When CNO-cycling at low temperatures was involved, the
star was deposited  at the blue end of the BHB. 
This disposition of progenitors along the BHB and evolution of BHB stars
to BL Her and W Vir stars is discussed further in Section 6.

The   anomalies common   among  W Vir stars but 
uncommon among BL Her stars 
are very unlikely to have a nucleosynthetic
origin. Anomalies include the
low [Al/Fe], 
[Ca/Fe], [Sc/Fe],  [Ti/Fe], and
[$s$/Fe] ratios. 
A catalog of explanations for the anomalies includes
 the standard invocation of errors in the
atmospheric parameters, non-LTE effects,  the inapplicability of standard
atmospheres invoking hydrostatic equilibrium to a low density atmosphere
subject to a pulsation, 
 non-LTE overionization
of atoms and ions by Lyman continuum photons from a shock wave in the pulsating
atmosphere,
 substantial helium enrichment
leading to a systematic overestimate of (say) [Fe/H] but to small(er) effects
on ratios such as [$s$/Fe],
and  reduction of the abundance of refractory elements by dust-gas separation in
the upper atmosphere or in a circumbinary disk.

Luck \& Bond (1989) showed that errors  in the atmospheric
parameters cannot erase the low [$s$/Fe] values, and by extension of 
their arguments, other  anomalous underabundances (relative to Fe)
cannot be attributed to injudicious choices of atmospheric
parameters. Non-LTE effects computed for standard atmospheres deserve
examination but are unlikely to account for example for differences
in abundance anomalies between and within the samples of
 BL Her and W Vir variables where stars of similar atmospheric
parameters can differ in the degree of their anomalies.

The possibility of Lyman continuum emission leading to departures from
LTE ionization equilibrium was mentioned by Barker et al. (1971)
following a suggestion by Wallerstein, and discussed in more detail
by Luck \& Bond (1989). Since the observable heavy elements from Y to Eu
are present in these atmospheres
 predominantly as singly-charged ions, all with ionization
potentials less than ionization potential of H, the idea has an
appeal.
 Emission in Balmer H$\alpha$ and the He\,{\sc i}
5876 \AA\ lines is a feature of the rising branch for the longer-period
Type II Cepheids for which $s$-process and other underabundances are
common.
 The fact that Eu does not show an abundance scatter like the $s$-process
elements is a strike against the proposal. It also
 does not account for the underabundance of
Al;  Al$^+$ has an ionization potential of 18.83 eV and Lyman emission at
18.8 eV must be weak and, certainly, Mg present in the atmosphere as Mg$^+$
with an ionization potential of 15.04 eV is never underabundant.
 The idea might be given a  stringent test
by observing a star at intervals throughout its pulsation and
seeing if the derived composition correlates with the presence of a shock as
revealed by line splitting. The weight of the evidence is that emission from
the shock wave is not a major source of the abundance anomalies. 
 
Addition of extensive amounts of $3\alpha$-processed and CN-cycled
material might foster the idea that the reported abundance anomalies are
a consequence of analysing the spectrum of a He-rich atmosphere on the
assumption that it has a normal He abundance. To first order, this
incorrect assumption results in [X/Fe] ratios close to the
actual values but a [Fe/H] that is overestimated.
 Luck \& Bond (1989)
speculated that a  low [$s$/Fe] ratio,
 a feature of normal stars of [Fe/H] $< -2$,
was being associated with a higher [Fe/H] because of an overlooked
helium enrichment.
The reader is referred to
their paper for arguments leading to the conjecture's rejection. Here, we
note their  point that the kinematics of the  W Vir stars
show that they belong to the
thick disk and not to the very metal-poor halo.  Yet, a thorough attempt to
measure the He/H ratio of the metal-rich variables would be welcomed.

Dust-gas separation is  evident in RV Tau variables (Giridhar
et al. 2005; Maas, Van Winckel, \& Waelkens 2002)
 except for the  metal-poor and cool variables. In oxygen-rich gas, the heavy
elements are among the first to be removed from the gas as
cooling occurs.
The demonstration by Alcock et al. (1998) that the RV Tau and Type
II Cepheids in the LMC define a single period-luminosity-color
relation suggests that the longer period Type II Cepheids in our sample
may be susceptible to  dust-gas separation. Earlier, we noted that several of
our stars show RV Tau-like photometric behavior and might be
promoted from W Vir to RV Tau status.
 The
signature of dust-gas
 separation, previously reported for ST Pup (Gonzalez \&
Wallerstein 1996),  is strikingly evident here for CC Lyr where
the spectrum contains lines of elements having a low condensation
temperature ($T_{\rm C}$),  i.e.,
lines of C\,{\sc i}, N\,{\sc i}, O\,{\sc i}, Na\,{\sc i}, S\,{\sc i}, and
Zn\,{\sc i}, but lines of higher condensation
temperature are poorly represented or not detected at all.
 Severe underabundances
are measured, e.g., [Mg/H] $\simeq$
[Fe/H] $\simeq -4$ but [S/H] $\simeq$ [Zn/H] $\simeq -0.5$. Figure 7
constructed for CC Lyr
shows the characteristic pattern for a strongly affected RV Tau variable;
the abundances [X/H] decline smoothly with increasing condensation
temperature.
The Mg and Fe deficiencies of CC Lyr are more extreme than any
reported previously for a RV Tau variable. Lines of the elements normally the
most depleted -- Al, Ca, Sc, and Ti, for example -- could not be detected
in our spectra.
 CC Lyr has a considerable
infrared excess from circumstellar dust. Its Fe deficiency is typical of
values reached in the warm post-AGB stars known to be spectroscopic binaries
(Van Winckel, Waelkens, \& Waters 1995).
One supposes that CC Lyr may also be a spectroscopic binary.
ST Pup, a W Vir star with a 
strong infrared excess and signatures of dust-gas separation,
 was shown by Gonzalez \& Wallerstein
 (1996) to be a spectroscopic
binary with an orbital period of 410 days.\footnote{The
dust-gas  signatures for ST Pup are rather ragged but this is possibly
attributable to larger than normal uncertainties in the derived
abundances resulting from the very few lines  available for most of
the elements:
[Fe/H] $= -1.47$ with [S/Fe] $= +1.29$ and [Zn/Fe] $= +1.41$ are
customary markers of severe dust-gas separation but [Al/Fe] $= +0.19$ is not.}

The signature of dust-gas separation seems present in other W Vir stars,
all of which show negative [Sc/Fe] values. Since
these stars are likely thick-disk members of different initial [Fe/H], we
correct the derived abundances [X/H] for the initial [X/H] of the
thick disk (Reddy et al. 2006) before presenting the $T_{\rm C}$ versus [X/H]
plots; all corrections are small.
 Figure 8 shows the [X/H] versus $T_{\rm C}$ plots for
CO Pup and W Vir.
In such cases, as in the examples of mildly affected RV Tau stars,
the scatter in [X/H] at a given condensation temperature may
be significant. This scatter is attributed to the understandable
failure of the condensation temperature to represent fully the
complex processes of dust formation and accretion of gas but not dust by
the star. 
 The decrease in [X/H] at the highest $T_{\rm C}$ is
clear for CO Pup. For W Vir, this signature of dust-gas separation is
critically dependent on the Sc abundance.
Here and in Figures 9 and 10, the $s$-process elements are
denoted by unfilled circles. Lack of a clear separation between the
unfilled and filled circles referring to elements of similar $T_{\rm C}$
show that the $s$-process products did not mix to the surface in the
time that these W Vir stars were on the AGB.
A clearer example of dust-gas separation
 is provided by V1711 Sgr with Al, Ti, and Sc, all 
underabundant by more than one dex relative to Fe (Figure 9). 
The signature of dust-gas separation also seems evident for SZ Mon, the
W Vir star with a RV Tau-like light curve and a strong infrared excess
(Figure 9). The evidence seems fair   for MZ Cyg but weaker for
RX Lib (Figure 10) where it relies on the low
Sc abundance. In all cases, the $s$-process elements are underabundant,
as expected when Al, Sc, and Ti are underabundant.
 Although several  key elements have not been measured for TW Cap, a halo star,
the indication is that dust-gas separation has not affected this star;
Ca, Ti, Sc, and $s$-process elements have the abundances expected of a 
star with [Fe/H] $-1.8$, a result anticipated from our previous reports on the
absence of dust-gas separation in intrinsically 
metal-poor RV Tau variables (Giridhar et al. 2005). 

Binarity has been advanced as a principal key to the dust-gas separation
exhibited by RV Tau stars;  Van Winckel (2003)
suggested that `binarity may very well be a common phenomenon among
RV Tau stars'. In the present sample, the known spectroscopic
binaries -- TX Del, IX Cas, and AU Peg --
are  not strongly affected by dust-gas separation but all are BL Her stars
for which dust-gas separation is not observed.
The affected W Vir star ST Pup is a binary (Gonzalez \& Wallerstein 1996).
Investigating the binary nature of the W Vir stars, not only those
affected by dust-gas separation, will call for intensive
radial velocity monitoring in order to resolve the orbital from the
pulsational velocity variations. It is noteworthy  that the
Fe-deficiency of CC Lyr is similar to that of the A-type
post-AGB stars, all known to be spectroscopic binaries 
(Van Winckel et al. 1995).

In short, the suggestion is clear: the principal reason for the various
abundance anomalies -- C, N, and O apart -- among the W Vir stars is that
their atmospheres have been partially cleansed of refractory elements.
The mass of the
convective envelope must be small to sustain a large deficiency of
refractory elements. Perhaps, not coincidentally a
 small envelope is also a condition required
to place the immediate progenitors of these variables
on the horizontal branch to the blue of the
red clump. There may also be a correlation between the magnitude of the
infrared excess and the visibility of the dust-gas separation. Stars
with a weak dust-gas separation are returning to a normal
composition following dissolution of circumstellar dust.  
The BL Her stars have yet to evolve to the upper AGB and to commence
dust production and, hence, absence of evidence of dust-gas separation
is  not surprising.

In discussing the abundance anomalies, we have overlooked 
the  possibility that a contributing factor
might be found in conditions existing in
the progenitors resident on the blue horizontal branch. 
Stars with the lowest ratios of envelope to core mass reside on 
the horizontal branch at effective temperatures ($T_{\rm eff} > 11500$ K)
such that severe atmospheric abundance anomalies are created by gravitational
settling and radiative levitation (Behr 2003), e.g., Fe overabundances
by a factor of 2 dex have been recorded. These stars are
likely to evolve along a track-bluenose. However, one supposes that such
anomalies are erased by the time that the stars have evolved off the 
horizontal branch and into the instability strip. This
supposition deserves observational and theoretical attention.

\section{Metallicities from photometry}

Accumulation and analysis of high-resolution spectra is a time
consuming process. A variety of statistical properties of the Type II Cepheids
may be more readily examined using a photometric estimate of metallicity
provided that  the estimate is reliable.
Here, we assess the reliability for three photometric systems
which have been applied to Type II Cepheids.

Harris (1981)
 used Washington photometry to obtain  the
metallicity [A/H] for a large sample of Type II Cepheids --
 see  Harris \& Wallerstein (1984) for two revisions. There are
13 stars not including CC Lyr in common with our sample. The comparison
of our [Fe/H] and [A/H] is shown in Figure 11. (The [A/H] for CC Lyr with its
highly non-standard composition is expected to depart appreciably from
the spectroscopic [Fe/H]: indeed, [A/H] of $-2.2$ versus [Fe/H] of about $-4$.)
It is clear from the figure that the photometric and spectroscopic
estimates are well correlated but offset such that [A/H] is overestimated
by about 0.5 dex. The offset is not dependent on the pulsation
period of the star.
The conclusion is that the Washington photometry provides, after a 0.5
dex downward revision,  metallicity estimates of satisfactory quality for
statistical purposes -- see, for example, the distribution functions
for metallicity of Type II Cepheid (Harris 1981).

Diethelm (1990) used Walraven {\it VBLUW} photometry to estimate the
metallicity of the shorter period Type II Cepheids. Five stars are
in common with our sample including a  RR Lyrae IK Hya (not discussed
here). Again, 
there is an offset between photometric and spectroscopic estimates:
[A/H] is overestimated by about 0.5 dex for the metal-rich stars
and possibly by 0.8 dex at low [Fe/H] (Figure 11). 

Str\"{o}mgren photometry was applied by Meakes, Wallerstein, \&
Opalko (1991). Excluding CC Lyr ([A/H] of $-2.3$), the  common stars
show that  [A/H] and [Fe/H] are  roughly correlated (Figure 11). 
 
In summary, the metallicities may be reliably provided by photometry
except for stars seriously affected by dust-gas separation.

\section{Concluding remarks}

Our sample of field Type II Cepheids is the first for which extensive
data on chemical compositions are provided. The study
complements prior modern analyses of $\kappa$ Pav, ST Pup, V553 Cen, and
RT TrA.  Our data on just two
halo (metal-poor) Type II
 Cepheids omit key elements -- e.g., C, N, and Na -- so that
no more than a perfunctory interpretation of these stars is offered.
Common to the BL Her and W Vir stars is evidence for addition to
their atmospheres of $3\alpha$-processed and CN-cycled material.
The distinguishing mark of the BL Her stars, the shorter period Type II
Cepheids, is their overabundance of Na, an overabundance not
detected among the W Vir stars. The distinguishing mark of the
W Vir stars, the longer period Type II Cepheids, is a composition
modified by dust-gas separation to varying degrees.

Our following discussion assumes that the post-BHB evolution
follows the lines given in Section 4.1 (Figure 2): evolution follows either a
track-direct or a track-bluenose and the order of development is
BHB $\rightarrow$ BL Her $\rightarrow$ W Vir $\rightarrow$ post-AGB.
Then, there  are two  evolutionary puzzles presented by these 
distinguishing marks:

1: BL Her stars are Na-rich and evolve to become W Vir stars but none of the analysed
W Vir stars with the possible exception of V1711 Sg are Na-rich.

2. W Vir stars have normal sodium abundances and have evolved from BL Her stars
but all of the analysed BL Her stars are Na-rich.

In the case of the first puzzle, the simplest explanation  is that a long-period
 (W Vir or RV Tau) variable resulting from a Na-rich BL Her star
transits the instability strip much faster than the W Vir star of normal
Na abundance. Our discussion of Na abundances of RV Tau stars showed that
BL Her-like Na overabundances are seen in approximately 10 per cent of
the sample.  If V1711 Sgr is Na-rich, a similar fraction of our
sample of ten W Vir  stars is Na-rich.  

Possible resolution of the second puzzle starts with the
conjecture  that changes to the sodium abundance are
determined prior to residence on the BHB by events
associated with the mass loss (and mixing) required to divert the
star from the red clump to the BHB. 
Sodium production by proton capture on $^{22}$Ne occurs in
association with H-burning by the CNO-cycles only at `high'
temperatures. Thus, the BHB may be
populated by Na-rich and Na-normal stars.
Stars deposited on the BHB without experiencing sodium enrichment
evolve to Na-normal  BL Her stars before  becoming Na-normal
W Vir stars.
 The fact that our sample of nine BL Her stars
does not include a Na-normal star
may be a consequence of  a lower production rate of Na-normal
to Na-rich BHB stars, and a more rapid crossing of the instability strip 
by the Na-normal 
than the Na-rich BL Her stars.  
Given that a star on a track-bluenose makes three crossings of the instability
strip at a period appropriate for a BL Her star, we identify the Na-rich
BL Her stars with these tracks. These tracks begin at the blue end of the
BHB. 
The stars on a track-direct start at the red end of the BHB, cross
the instability strip once as a BL Her star and then evolve up the AGB to
enjoy thermal pulses before entering the instability strip as a W Vir or
RV Tau star. Such BHB stars are presumed to lack a sodium enrichment.
Examination of a larger sample of BL Her stars is, then, expected to
provide an example or two of a Na-rich variable.

In seeking to  resolve puzzles in stellar evolution,
 one endeavors to connect the stars with their progenitors, 
descendants, and close relatives among field and cluster stars.
 The vast majority of our sample
are probably thick disk citizens. Close relatives are the
RR Lyrae stars at one extreme and RV Tau and post-AGB stars at the other
extreme. We have already noted similarities between the compositions of the
W Vir and RV Tau stars.
 Progenitors of the BL Her and W Vir stars are  on the blue
horizontal branch. Identified BHB stars have halo kinematics
and metallicities and are surely progenitors of very metal-poor variables
such as UY Eri and TW Cap. Thick disk BHB stars have proven
elusive: `There are, however, {\it disk} RR Lyrae stars in the nearby
field, but there are (as far as we know) no corresponding field BHB stars that
have disk kinematics' (Kinman et al. 2000). It would be extremely
valuable to locate some thick disk BHB stars and to
determine their compositions, especially their Na abundances. 
Our attribution of BL Her stars to a track-bluenose and W Vir stars to a
track-direct implies the redder BHB stars should have a normal
Na abundance and the bluer BHB stars should be sodium rich.

 Fresh insights into the phenomenon of the Type II Cepheids may
be gleaned from abundance analyses of these variables in
globular clusters. Published results apply to W Vir or RV Tau stars
(i.e., periods greater than 10 days) in four
globular clusters and five post-HB stars including three variables
in $\omega$ Cen (cf. Wallerstein's [2002] review).
All variables bar one in $\omega$ Cen have
a period representative of W Vir stars and the exception with a
period of 4.5 days is likely a BL Her star.
Results for the globular clusters M2, M5, M10, M12, and M28 -- one
from each cluster with the exception of two from M5 -- are provided
variously by Gonzalez \&amp; Lambert (1997), Carney et al. (1998), and
Klochkova et al. (2003). Gonzalez \&amp; Wallerstein (1994) give
abundances for the $\omega$ Cen stars. The clusters have metallicities
from [Fe/H] of $-1.2$ to $-1.7$ with the stars from $\omega$ Cen
spanning the range from $-1.7$ to $-2.1$. Thus, the sample is
collectively more metal-poor than the large majority of the variables
discussed in this paper.

Perhaps, the most significant conclusion to be drawn from these
metal-poor stars is that there is no evidence that their atmospheres
have been affected by dust-gas separation, a result expected from
observations of metal-poor field RV Tau stars (Giridhar et al. 2005).
Interpretation of light element abundances (C to Al) is complicated
by the fact that many clusters show star-to-star correlated
abundance variations that are entirely missing from samples of field stars.
Since such variations are present among main sequence members of a cluster,
they cannot be explained wholly as internal processing and mixing.
In this respect, Carney et al. (1998) show that the variable V42 in M5
has the  Na enrichment and O deficiency seen in the most extreme Na-rich and
O-poor red giants, i.e., the Na and O abundances of V42 may be unrelated to
its transition from the red giant to the horizontal branch.
Notwithstanding this novel complication presented by the clusters and
their variables, the preliminary forays into abundance analyses of
cluster Type II Cepheids and their horizontal branch progenitors
deserve to be extended: for example, the key elements of C, N, and O
remain almost untouched by the published analyses (see Gonzalez \&amp; Wallerstein
for C, N, and O among $\omega$ Cen stars, and Klochkova et al. (2003) for
C, N, and O for V1 in M12).\footnote{A  remarkable result concerns the
Al abundance in the $\omega$ Cen stars. In four of the five stars, Al was
underabundant by $-0.3$ to $-1.1$ as measured by [Al/Fe] but without the
attendant Ca and Sc deficiencies expected from dust-gas separation. Yet more
remarkable is the observation that in the fifth star, Gonzalez \&amp; Wallerstein
(1994) found a strong Al overabundance ([Al/Fe] = $+1.2$) in an O-deficient
star with a normal Na abundance.} Extension of the work on cluster
variables to the BL Her stars would also be of interest. This would be
of special interest if these BL Her stars are shown to be Na-rich. Material
rich in Na from proton capture on $^{22}$Ne is possibly also He-rich.
The pulsational properties of Type II Cepheids are dependent on the
He abundance as well as mass and metallicity (see Bono et al. [1997] for
references to these dependencies). Since BL Her stars in globular
clusters have an accurately determined absolute luminosity in
contrast to field stars where the luminosity is never well
determined, it may be possible to determine their helium
abundance.

Immediate   descendants of the Type II Cepheids are red giants and
 must be picked out
from the (presumably) more numerous giant stars evolved from clump
giants. Anomalous compositions are a likely indicator.
Wallerstein \& Gonzalez (1996) and Wallerstein et al. (2000) noted the
similarities between the compositions of the C-rich Type II Cepheids
and the R-type carbon stars (Dominy 1984) where products of the
$3\alpha$-process are greatly in evidence. At the other extreme, IX Cas and
the Weak G-band giants may be related examples of stars where the stellar
atmosphere is dominated by CNO-cycled products and $3\alpha$--processed
products are absent or just a minor contaminant. Although intriguing, these
suggested links between Type II Cepheids and peculiar red giants do not
directly suggest an interpretation  in terms of stellar evolution. 
The observers' {\it Deus ex machina} for linking a Type II
Cepheid to a peculiar red giant  may be the He-core flash but,
as noted earlier, theoreticians discount this possibility. 
Perhaps,  the appeal should be  not to an episode in the life of a single star
but to one  in the evolution of a binary
star.  Such an  appeal, however, may be a forlorn one in the case of the R-type
carbon stars and Weak G-band giants because neither class
seem to have an unusual degree of binarity (Tomkin, Sneden, \&
Cottrell 1984; McClure 1985).

We thank Ron Wilhelm for helpful correspondence.
This research has made use of NASA's Astrophysics Data System and the
Centre de Don\'{e}es de Strasbourg's  SIMBAD
astronomical database.
 We also thank the anonymous referee for valuable suggestions.
This research has been supported in part by the Robert A. Welch Foundation of
Houston, Texas.

\begin{figure}
\plotone{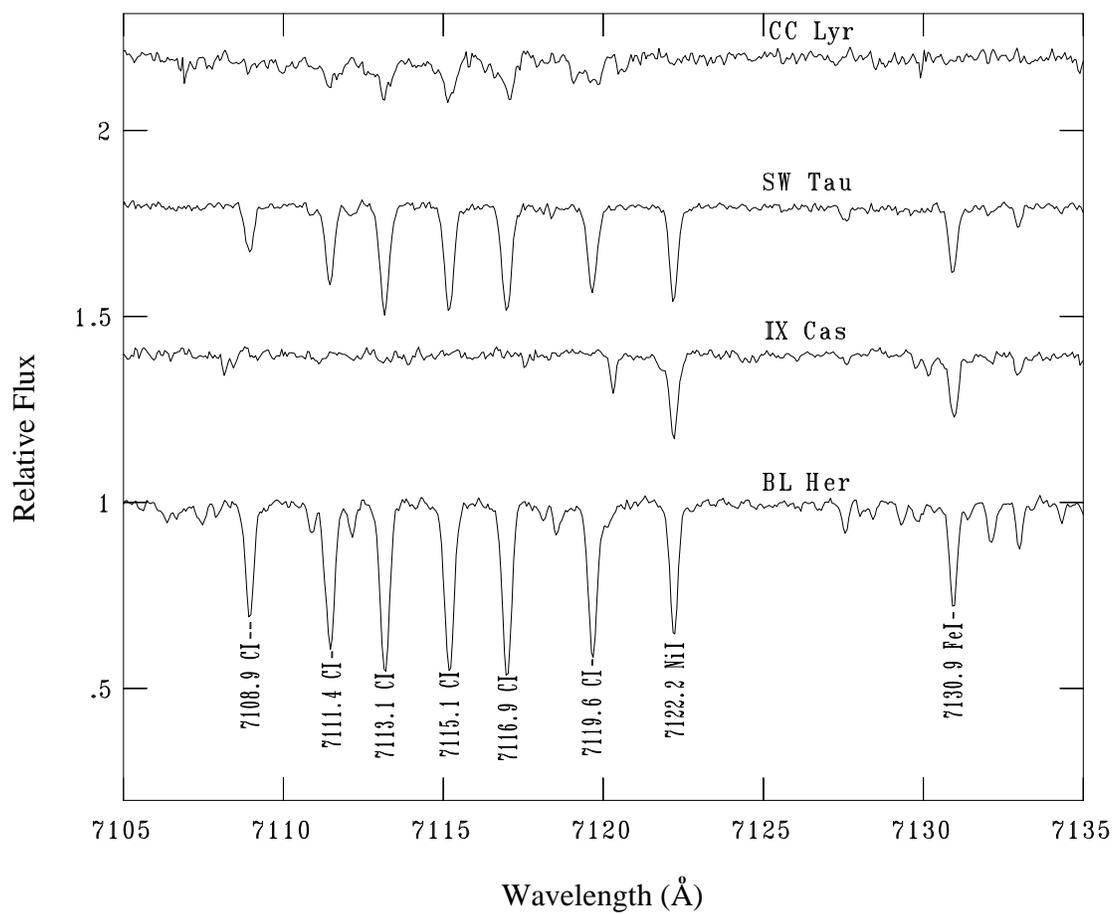}
\figcaption{The interval 7105--7135 \AA\ for CC Lyr, SW Tau, IX Cas, and BL Her.
Note the 
absence of the C\,{\sc i} lines for IX Cas, and the absence of the Fe\,{\sc i}
and Ni\,{\sc i} lines for CC Lyr.}
\end{figure}


\begin{figure}
\figcaption{A HR diagram showing the location of the Population II instability
strip and approximate lines of constant period. Superimposed  are
two evolutionary tracks for two stars: the track for the  0.6$M_\odot$ star is
a track-direct and that for the 0.536$M_\odot$ is a track-bluenose (see text).
The location of the RR Lyr variables and the approximate positions of
RV Tau variables are marked. (After Gingold 1985)}
\end{figure}

\begin{figure}
\plotone{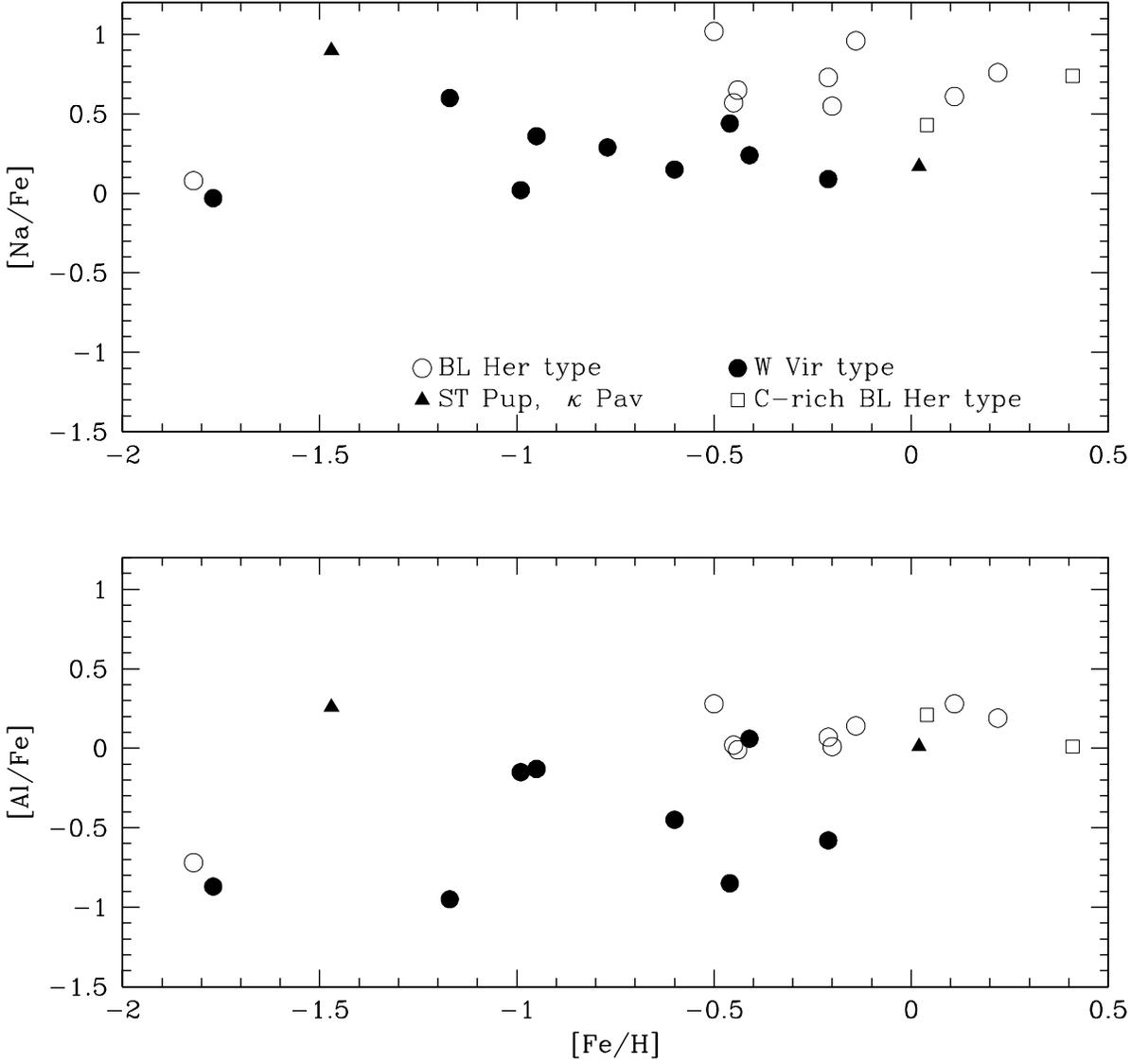}
\figcaption{The [Na/Fe] (top panel) and [Al/Fe] (bottom panel) ratios
versus [Fe/H]. The key to the symbols is given in the top panel.
Symbols: unfilled circles -- Our BL Her stars,
 filled
circles -- Our W Vir stars, unfilled squares -- The C-rich BL Her
stars RT TrA and V553 Cen, filled triangles-- The W Vir stars $\kappa$ Pav at
[Fe/H] $\simeq 0.0$ and ST Pup at [Fe/H] $\simeq -1.5$.} 
\end{figure}

\begin{figure}
\plotone{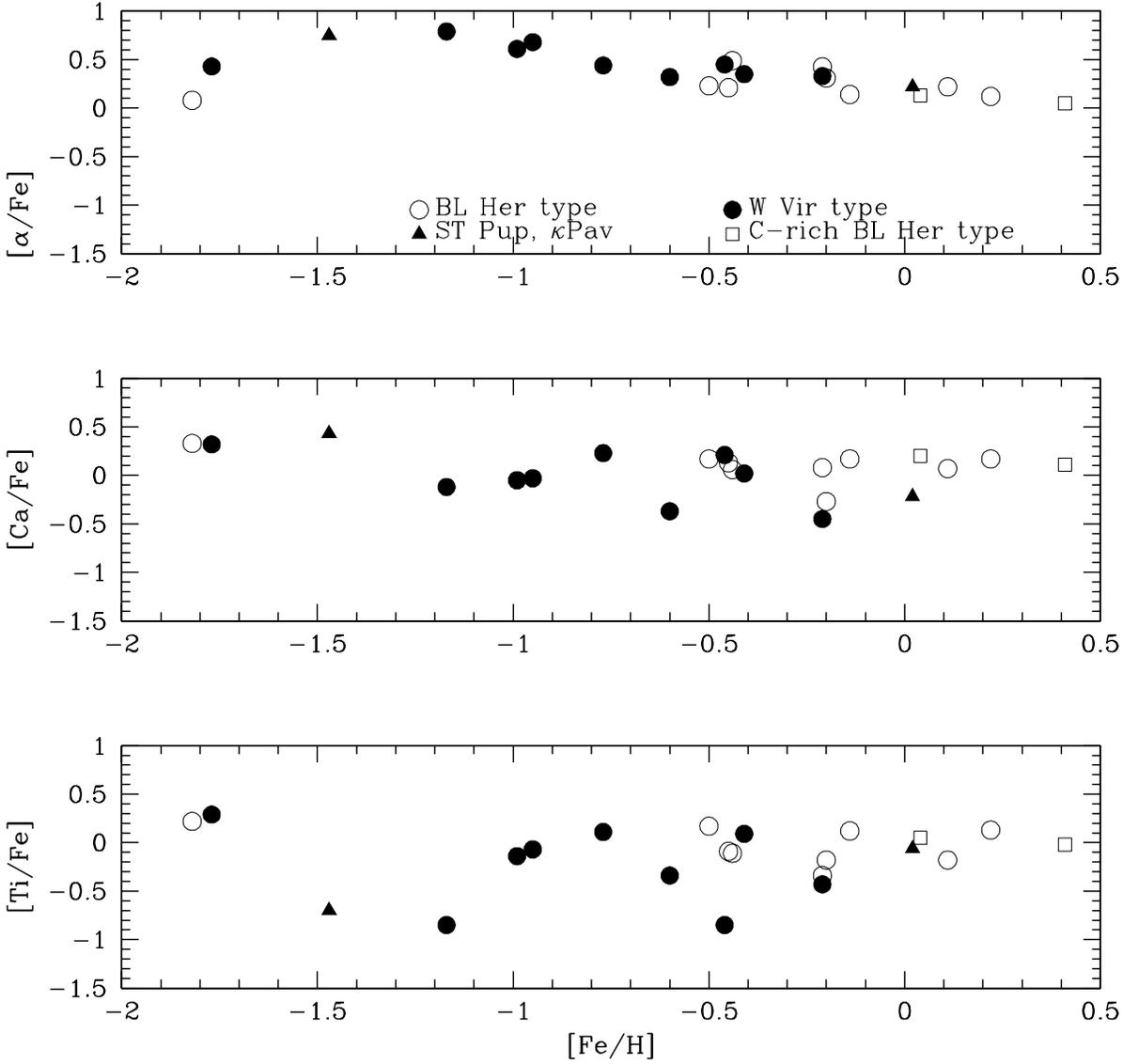}
\figcaption{The run of the abundances of $\alpha$-elements with [Fe/H].
The top panel shows the ratio [$\alpha$/Fe] versus [Fe/H] where
[$\alpha$/Fe] is the mean of the ratios [Mg/Fe], [Si/Fe], and [S/Fe].
The middle panel shows [Ca/Fe] versus [Fe/H]. The bottom panel
shows [Ti/Fe] versus [Fe/H]. Symbols are as in Figure 3. 
}
\end{figure}

\begin{figure}
\plotone{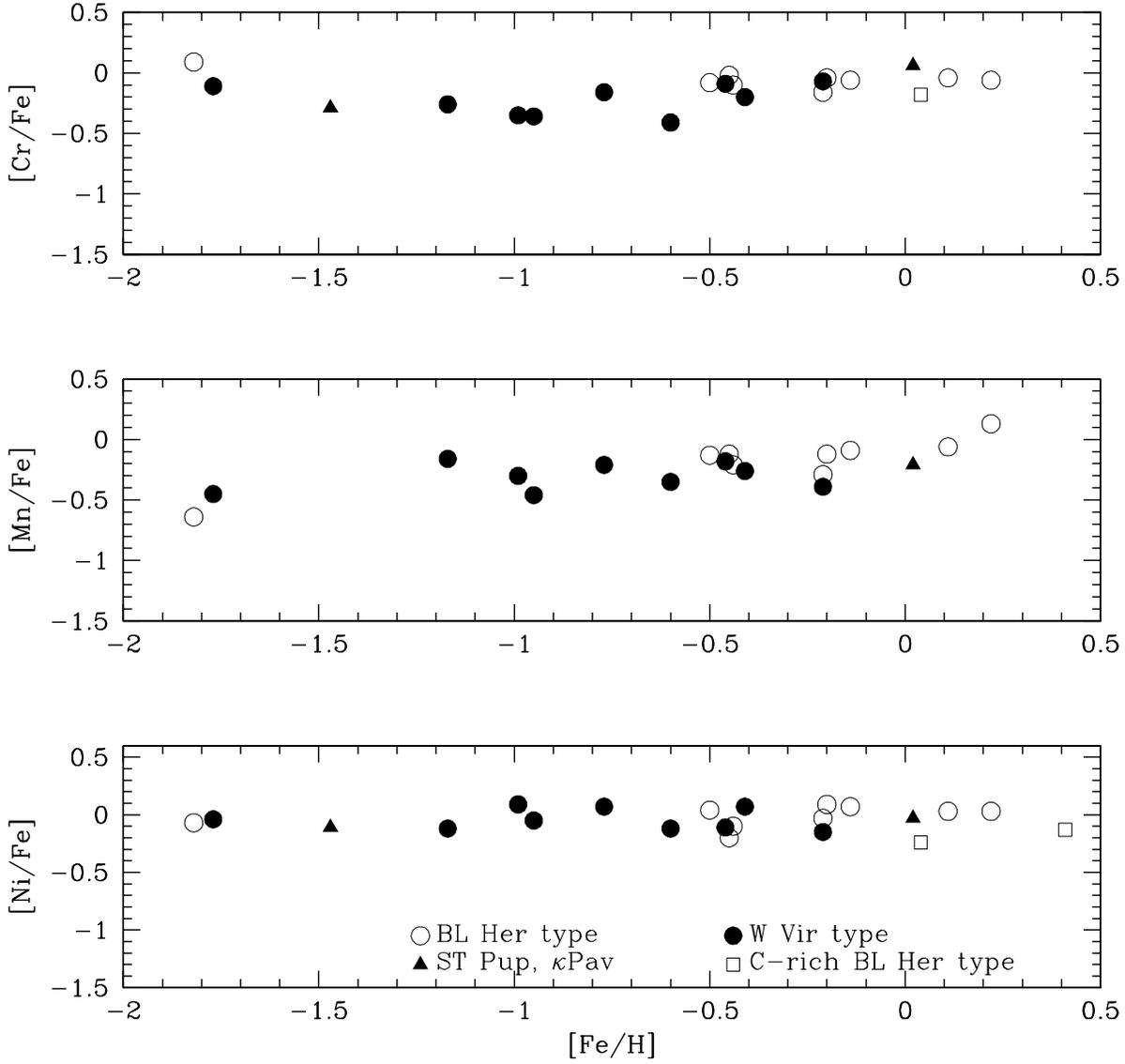}
\figcaption{The run of the ratios [Cr/Fe] (top panel), [Mn/Fe] (middle
panel), and [Ni/Fe] (bottom panel) versus [Fe/H]. Symbols are as in
Figure 3.}
\end{figure}

\begin{figure}
\plotone{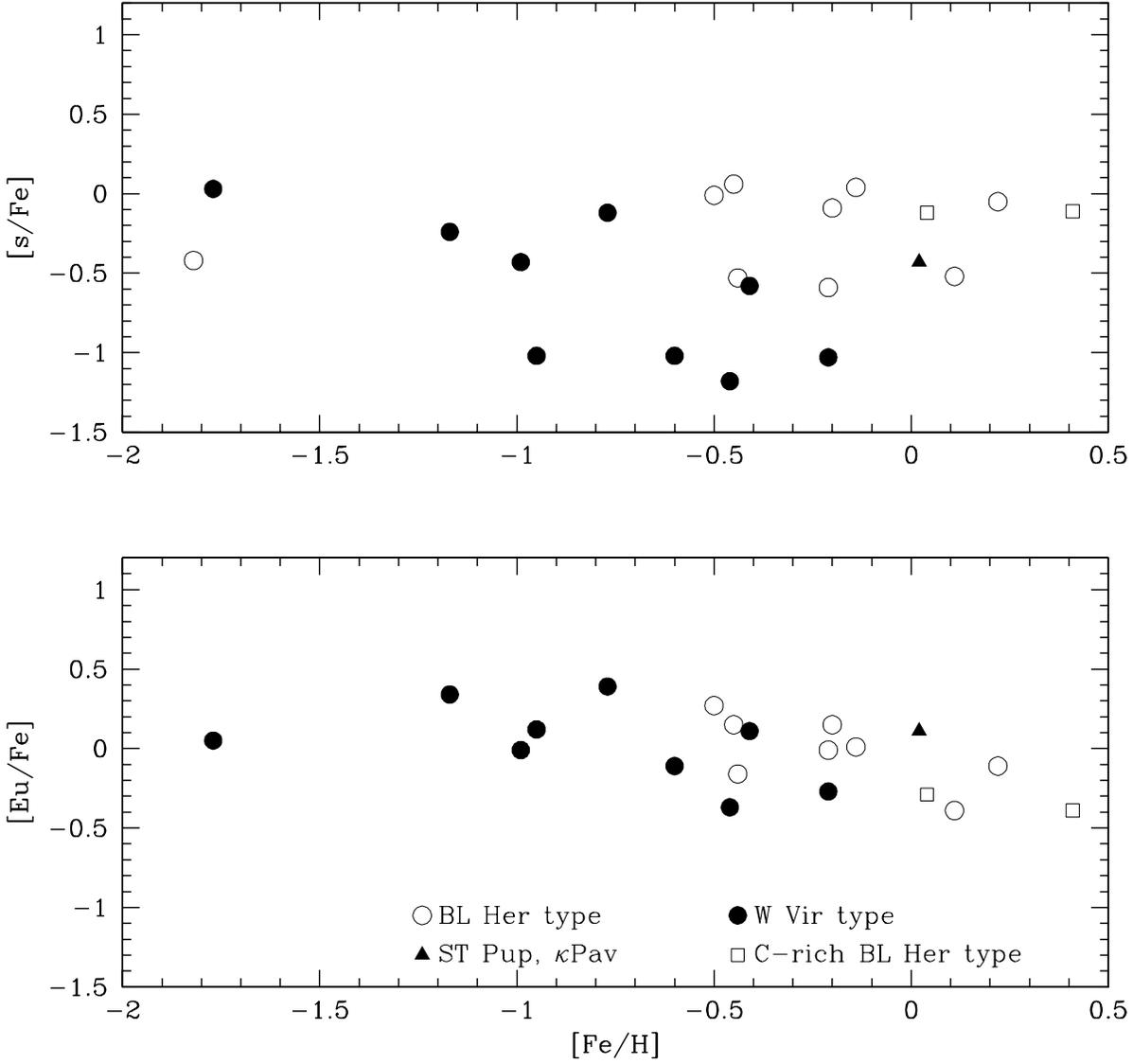}
\figcaption{The run of the ratio [$s$/Fe] (top panel) and [Eu/Fe]
(bottom panel) versus [Fe/H]. See text for the elements
contributing to the mean [$s$/Fe].  Symbols are as in Figure 3.}
\end{figure}

\begin{figure}
\plotone{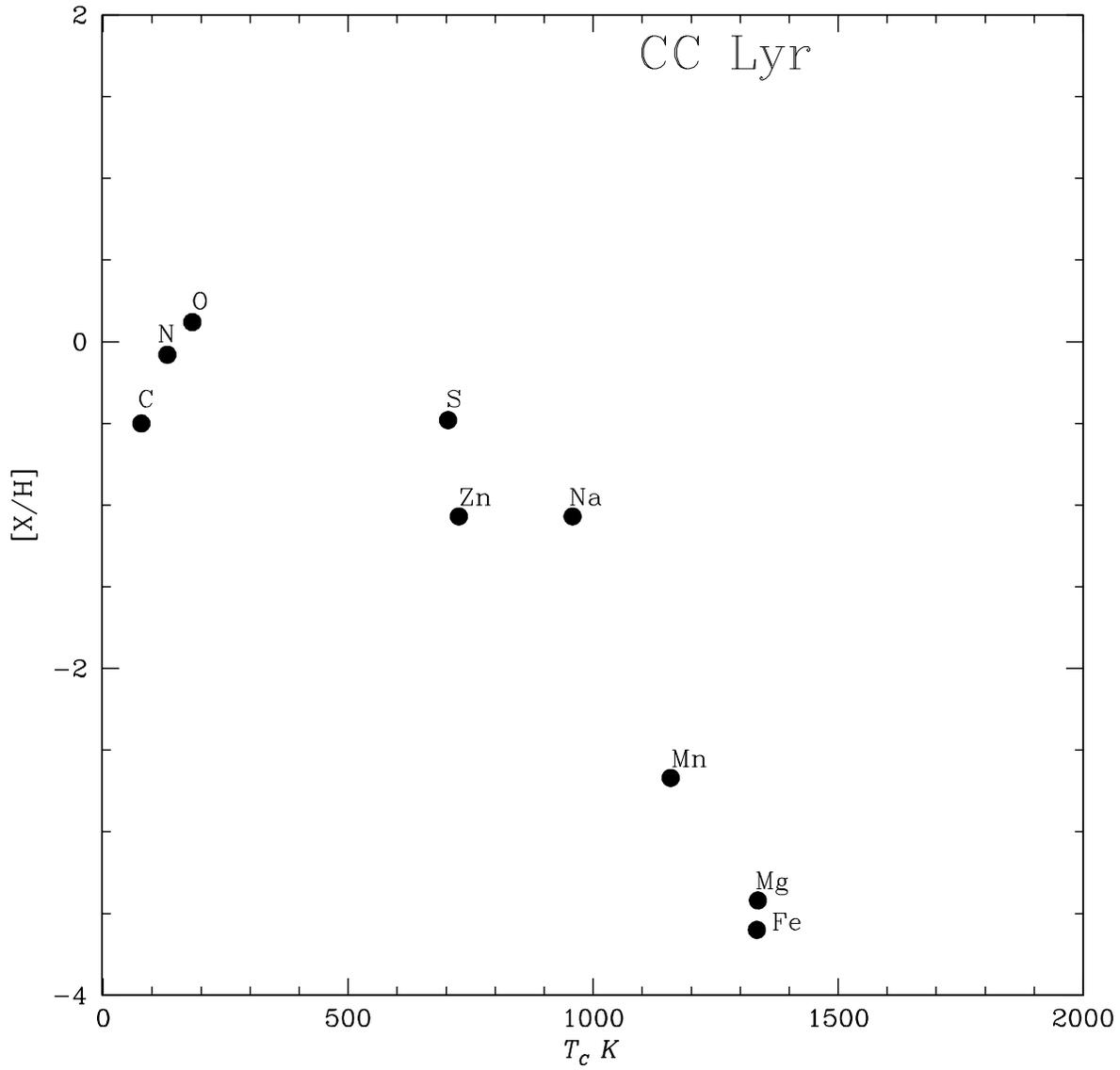}
\figcaption{Abundances [X/H] versus condensation temperature $T_C$
for CC Lyr. Elements are identified by their chemical symbol.}
\end{figure}

\begin{figure}
\plotone{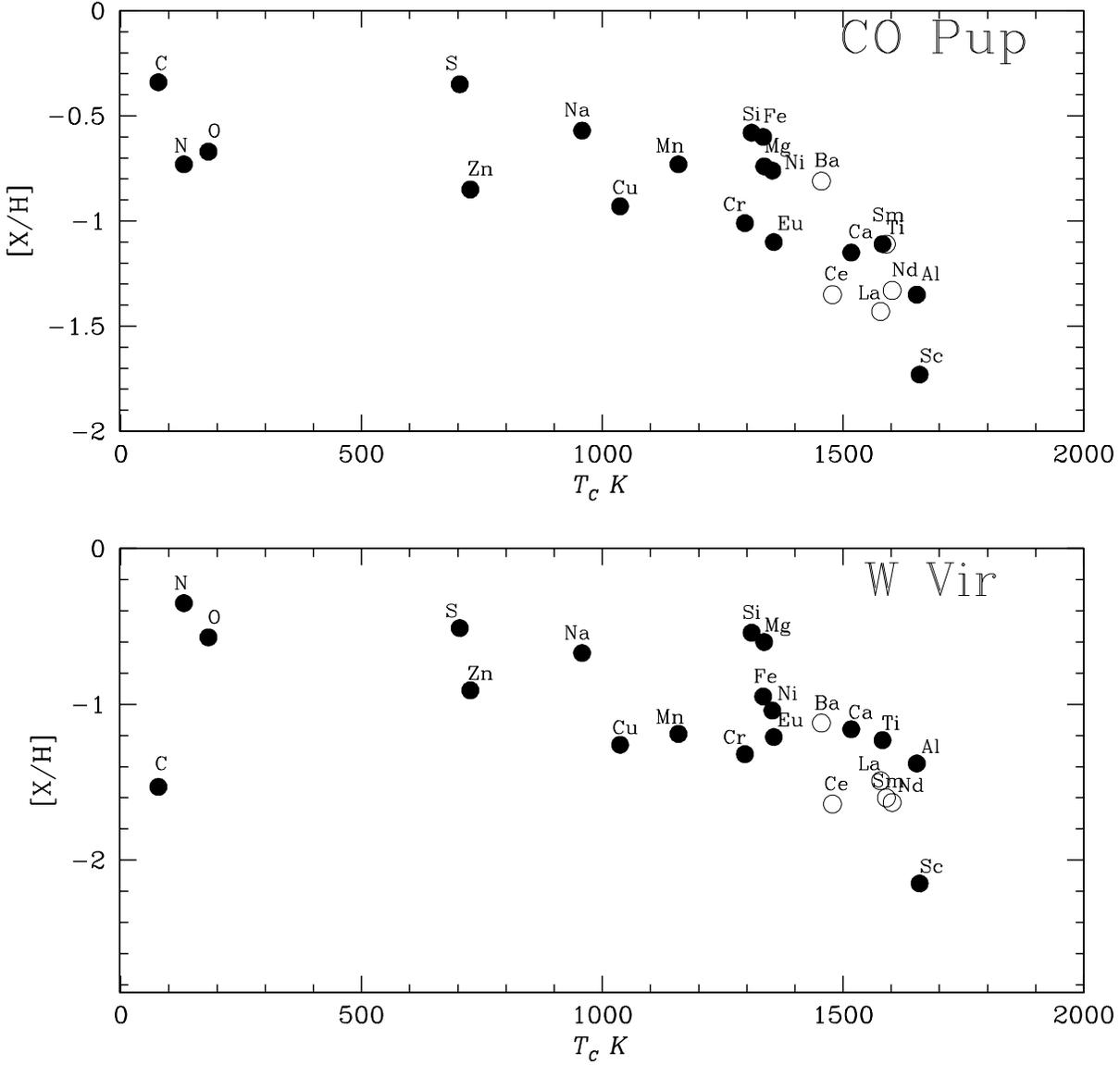}
\figcaption{Adjusted abundances (see text) [X/H] for CO Pup (top
panel) and W Vir (bottom panel) versus
condensation temperature $T_C$. Elements are identified by their
chemical symbol. Unfilled circles refer to elements made principally
by the $s$-process whose abundance might have been enhanced by thermal
pulses on the AGB.}
\end{figure}

\begin{figure}
\plotone{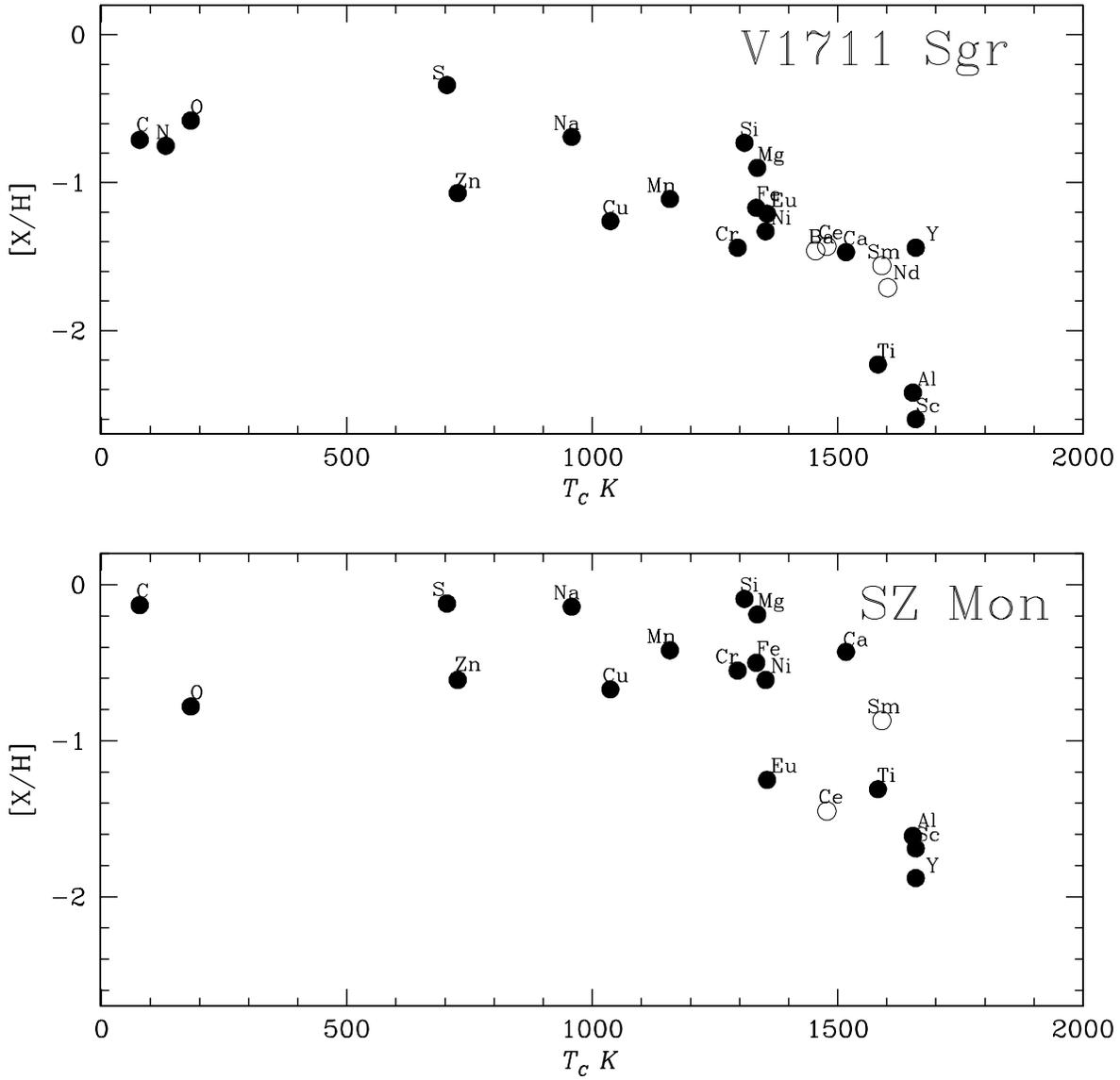}
\figcaption{Adjusted abundances (see text) [X/H] for
V1711 Sgr (top panel) and SZ Mon (bottom panel) versus
condensation temperature $T_C$. Elements are identified by their
chemical symbol.}
\end{figure}

\begin{figure}
\plotone{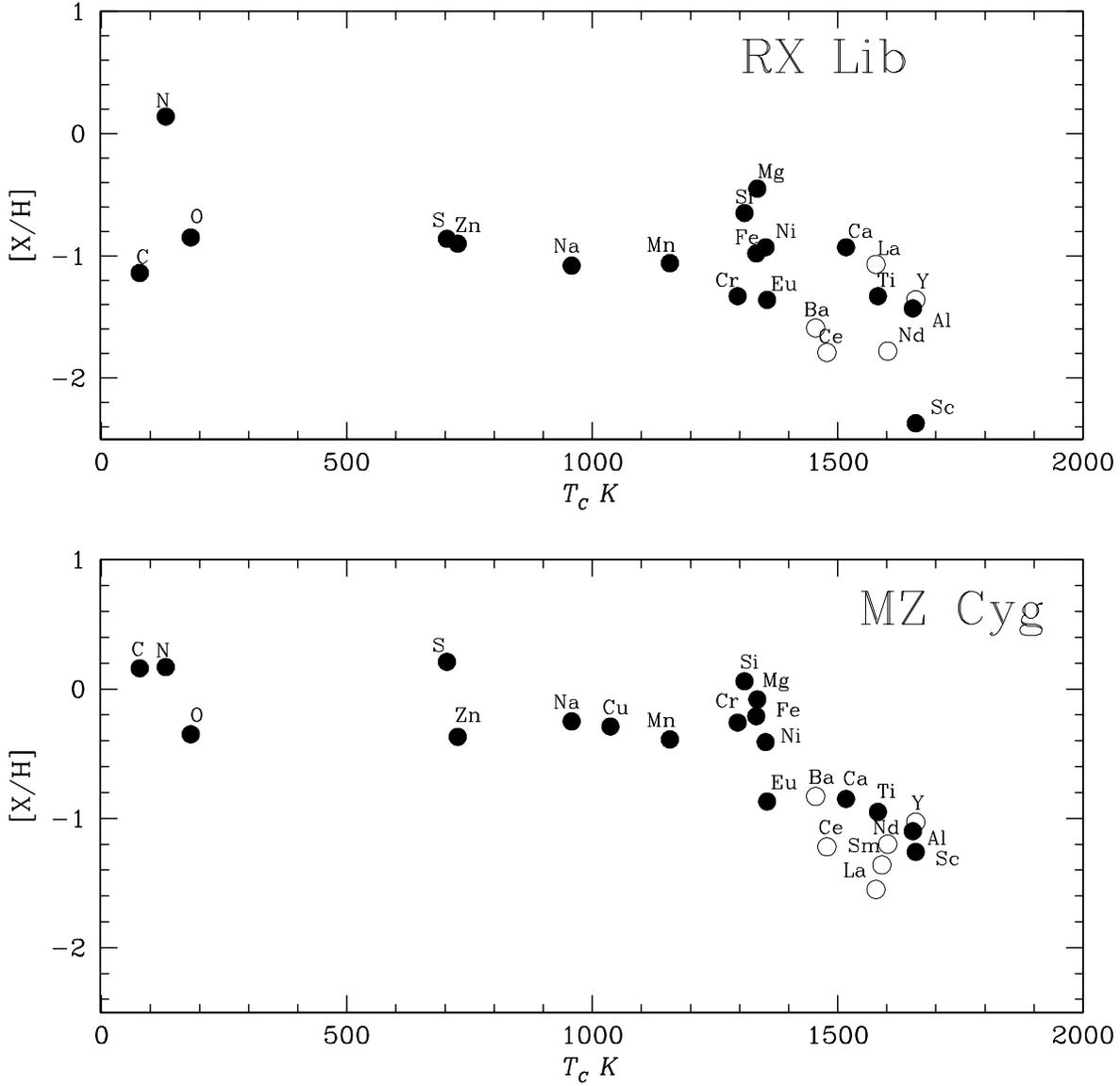}
\figcaption{Adjusted abundances [X/H] for RX Lib (top panel)
 and MZ Cyg (bottom panel) versus
condensation temperature $T_C$. Elements are identified by their
chemical symbol.}
\end{figure}

\begin{figure}
\plotone{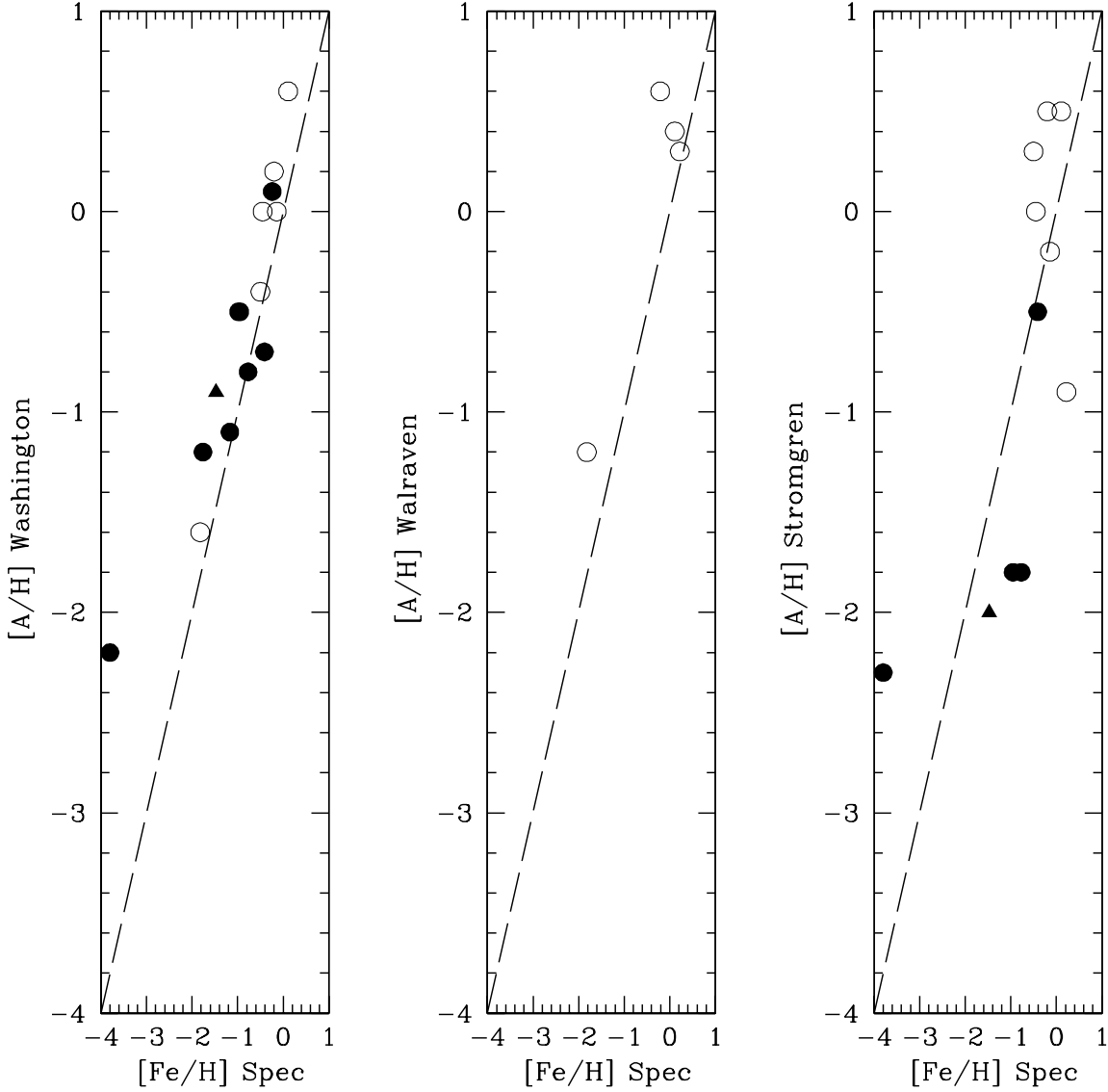}
\figcaption{Comparisons of our spectroscopic and published
photometric metallicities. The lefthand panel shows the comparison
with the results from Washington photometry (Harris 1985),
the middle panel with  Walraven
photometry (Diethelm 1990), and the righthand panel with Str\"{o}mgren
photometry (Meakes et al. 1991). The dashed line in each panel
indicates equality between spectroscopic and photometric metallicity. The key to the symbols is given in Figure 3.}
\end{figure}

\setcounter{table}{0}
\begin{table}[htdp]
\caption{STELLAR PARAMETERS FROM THE Fe-LINE ANALYSIS}
\footnotesize
\begin{center}
\begin{tabular}{lrllcrrllrrr}
\hline
\hline
Star & Period & UT Date & Model  &
$\xi^{b}_{t}$ & Fe\,{\sc i}$^{c}$ & && \ \ \ \ \ \ \ Fe\,{\sc ii}$^{c}$\\
\cline{4-4}\cline{6-7}\cline{9-10}
& (days) & & T$_{\rm eff}$,  log g, [Fe/H]$^{a}$ & (km s$^{-1}$) &
   log $\epsilon$ & $n$& & log $\epsilon$ & $n$\\
\hline
BX Del & 1.1 & 2005 Jul 3 & 6250, 1.0, -0.2 & 3.0 & 7.28 $\pm$ 0.16 &
99 & & 7.19 $\pm$ 0.19 & 24 \\
VY Pyx & 1.2 & 2005 Jan 1 & 5750, 1.5, -0.4 & 3.0  & 7.05 $\pm$ 0.13 & 76 & &
6.97 $\pm$ 0.14 & 12\\
BL Her & 1.3 & 2005 May 3 & 6500, 2.0, -0.1  & 2.5  & 7.32 $\pm$ 0.11 &
122 & & 7.30 $\pm$ 0.13 & 38\\
SW Tau & 1.6 & 2004 Oct 8  & 6250, 2.0, +0.2 & 3.0 & 7.69 $\pm$ 0.12 &
115 & & 7.66 $\pm$ 0.19 & 24 \\
UY Eri & 2.2  & 2004 Dec 18 & 6000, 1.5, -1.8 & 2.9  & 5.66 $\pm$
0.15 & 79 & & 5.59 $\pm$  0.10 & 17 \\
AU Peg & 2.4 & 2005 Sept 24 & 5750, 1.5, -0.2 & 5.3 & 7.26 $\pm$ 0.16
& 37 & & 7.24 $\pm$ 0.15 & 11\\
DQ And & 3.2 & 2004 Dec 19 & 5500, 1.5, -0.5 & 3.0  & 7.00 $\pm$
0.16 & 91 & & 7.00 $\pm$ 0.13 & 15\\
TX Del & 6.2 & 2004 Aug 9 & 5500, 0.5, +0.1 & 3.7 & 7.56  $\pm$  0.09
& 59 & & 7.56 $\pm$  0.17 & 16\\
IX Cas & 9.2 & 2004 Oct 09 & 6250, 1.0,  -0.4 & 3.3 & 6.97 $\pm$  0.11
& 108 & & 6.93 $\pm$ 0.14 & 35\\
AL Vir & 10.3 & 2004 May 3 & 5500, 1.0, -0.4 & 3.5  & 7.07 $\pm$
0.09 & 85 & & 7.04 $\pm$  0.13 & 27\\
                &       & 2005 Jan 30 & 6500, 1.5, -0.4 & 3.2 & 7.04
$\pm$  0.10 & 92 & & 7.01 $\pm$  0.12 & 24\\
AP Her & 10.4 & 2004 Nov  8 & 6500, 1.0, -0.7 & 2.7 & 6.70 $\pm$
0.11 & 103 & & 6.65 $\pm$ 0.17 & 36\\
CO Pup & 16.0 & 2004 Dec 19 & 5000, 0.5, -0.6 & 4.3 & 6.81 $\pm$ 0.14 &
130 & & 6.88 $\pm$ 0.16 & 12\\
SZ Mon & 16.3 & 2005 Jan 30 & 4700, 0.0, -0.4 & 3.7 & 7.02 $\pm$ 0.16 &
59 & & 6.95 $\pm$  0.09 & 8\\
W Vir & 17.3 & 2005 Dec 18  & 5000, 0.0, -1.0 & 3.8 & 6.48 $\pm$  0.13 &
110 & &6.53 $\pm$ 0.13 & 21\\
MZ Cyg & 21.4 & 2005 Jul 03 & 4750, 0.5, -0.2 & 4.7 & 7.21 $\pm$ 0.16
& 56 & & 7.27 $\pm$ 0.18 & 8\\
CC Lyr & 24.2 & 2004 Aug 9 & 6250, 1.0, -3.5 & 3.5 & 3.83 $\pm$ 0.04
& 2 & & 3.73 & 1\\
              & & 2005 Jul 3 & 6250, 1.0, -3.3 & 4.5 & 3.32 $\pm$ 0.04 &
2 & & . . . &  . . \\
RX Lib & 24.9 & 2005 Apr 24  & 5250, 0.0, -1.0  & 2.9 & 6.50 $\pm$ 0.17 & 61
& & 6.43 $\pm$ 0.15 & 14\\
TW Cap & 28.6 & 2004 Dec 8  & 5250, 0.5, -1.8 & 3.1 & 5.70 $\pm$
0.13 & 91 & & 5.73 $\pm$  0.14 & 31\\
      &           & 2005 Jul 3 & 6000, 0.0, -1.8 & 3.8 & 5.66 $\pm$ 0.14 &
33 & & 5.64 $\pm$ 0.18 & 13\\
V1711 Sgr  & 28.6 & 2005 Jul 3  & 5000, 0.5, -1.2  & 4.1 & 6.28 $\pm$ 0.16 & 67
& & 6.28 $\pm$ 0.17 & 15\\
\hline
\end{tabular}
\end{center}
$^{a}$ The value of T$_{\rm eff}$ is in kelvins; log g is in cgs,
[Fe/H] in dex.

$^{b}$ The symbol $\xi_{t}$ represents the microturbulence determined
from the Fe\,{\sc i} lines.

$^{c}$ The column headed log $\epsilon$ gives the mean abundance
relative to H (with log $\epsilon_{\rm H}$ = 12.00).  The standard
deviations of the means, as calculated from the line-to-line scatter,
are given.  The quantity $n$ is the number of considered lines.
\label{default}
\end{table}

\setlength{\oddsidemargin}{-1.5cm}
\setlength{\evensidemargin}{-.5cm}
\setlength{\textwidth}{11in}

\setcounter{table}{1}

\begin{table}
\caption{ELEMENTAL ABUNDANCES - THE BL HER VARIABLES}\vspace{.25cm}
\tiny
\begin{tabular}{lcc@{\hspace{-.5mm}}rrrrrrrr@{\hspace{-.005mm}}rrrrrrr@{\hspace{-.005mm}}rrrrr}

\hline
\hline
&&&&&&&&&&&&&&&&&&&&&&\\*[-.30cm]
& &  BX Del$^{b}$&& &  VY Pyx &&&BL Her&&&SW Tau &&& UY Eri &&& AU 
Peg &&& DQ And\\
\cline{3-4}\cline{6-7}\cline{9-10} \cline{12-13} \cline{15-16} 
\cline{18-19} \cline{21-22}

&&&&&&&&&&&&&&&&&&&&&&\\*[-.30cm]
Species & $\log \epsilon_{\odot}^{a}$&  [X/Fe] &  n &  &    [X/Fe] & 
n  &   &   [X/Fe] &  n  &  &    [X/Fe] &  n  &  &     [X/Fe] &  n  & 
&    [X/Fe] &  n  &  &    [X/Fe] &  n\\
\hline
&&&&&&&&&&&&&&&&&&&&&&\\*[-.30cm]
C\,{\sc i} & 8.39 & -0.02 & 14 & & -0.24 & 6 & & +0.44 & 29 & & 
+0.77 & 22 & &  \nodata & \nodata & &+0.07 & 5 &&  +0.08 & 7 \\
N\,{\sc i} & 7.78 & +1.60 & 3 & & +1.51 & 3 & & +1.09 & 6 & & +1.26 & 
10 & & $<$-0.1 & 3 & & +1.01 & 4 & & +0.82 & 4 \\
O\,{\sc i} & 8.66 & +0.41 & 2 & & +0.51 & 3 & & +0.23 & 5 & & -0.03 & 
6 & & +0.52 & 2 & & +0.35 & 3 & & +0.18 & 3\\
Na\,{\sc i} & 6.17 & +0.73 & 3 & & +0.65 & 3 & & +0.96 & 3 & & +0.76 
& 3 & & +0.08 & 1 & &+0.55 & 2 & & +0.57 & 4\\
Mg\,{\sc i}  & 7.53 & +0.36 & 4 & & +0.47 & 4 & & +0.12 & 1 & & +0.02 
& 1 & & +0.29 & 4 && +0.15 & 3 & & +0.11 & 4\\
Al\,{\sc i} & 6.37 & +0.07 & 5 & & -0.01 & 5 & & +0.14 & 3 & & +0.19 
& 4 & & -0.72 & 2 & & +0.01 & 1 & & +0.02 & 4 \\
Si\,{\sc i} & 7.51 & +0.48 & 9 & & +0.33 & 11 & & +0.29 & 18 & & 
+0.28 & 17 & & -0.34 & 1 & & +0.29 & 6 & & +0.20 & 8 \\
Si\,{\sc ii} & 7.51 & \nodata & \nodata & & +0.61 & 2 & & +0.01 & 3 & 
& \nodata & \nodata & & +0.06 & 2 & & \nodata & \nodata & & +0.23 & 
2\\
S\,{\sc i}  & 7.14 & +0.44 & 6 & & +0.54 & 4 & & +0.16 & 6 & & +0.06 
& 6 & & \nodata & \nodata & & +0.52 & 3 & & + 0.32 & 3\\
Ca\,{\sc i} & 6.31 & +0.08 & 9 & & +0.06 & 5 & & +0.17 & 19 & & +0.17 
& 18 & & +0.33 & 11 & & -0.27 & 4 & & +0.13 & 7 \\
Sc\,{\sc ii} & 3.05 & -0.45 & 4 & & 0.00 & 5 & & +0.08 & 7 & & +0.22 
& 8 && 0.00 & 11 & & +0.07 & 3 & & +0.10 & 6\\
Ti\,{\sc i}  & 4.90 & -0.38 & 2 & &-0.07 & 2 & & +0.14 & 1 & & +0.17 
& 3 & & \nodata & \nodata & & -0.15 & 4 & & -0.03 & 4 \\
Ti\,{\sc ii} & 4.90 & -0.30 & 6 & & -0.16 & 6 & & +0.11 & 29 & &
+0.10 & 17 & & +0.22 & 15 & & -0.21 & 2 & & -0.16 & 6\\
Cr\,{\sc i}  & 5.64 & -0.24 & 8 & & -0.14 & 6 & & -0.04 & 15 & & 
-0.05 & 13 & & +0.10 & 6 & & -0.15 & 3 & & -0.06 & 8\\
Cr\,{\sc ii} & 5.64 & -0.08 & 6 & & -0.06 & 7 & & -0.07 & 17 & & 
-0.08 & 18 & & +0.08 & 4 & & +0.06 & 4 & & +0.02 & 2\\
Mn\,{\sc i} & 5.39 & -0.29 & 3 & & -0.21 & 5 & & -0.09 & 8 & & +0.13 
& 9 & & -0.64 & 3 & & -0.12 & 2 & & -0.12 & 4\\
Ni\,{\sc i} & 6.23 & -0.03 & 12 & & -0.10 & 13 & & +0.07 & 44 & & 
+0.03 & 45 & & -0.07 &  2 & & +0.09 & 8 & & -0.20 & 20 \\
Cu\,{\sc i} & 4.21 & +0.19 & 1 & & +0.31 & 1 & & +0.21 & 1 & & +0.38 
& 1 & & \nodata & \nodata & & \nodata & \nodata & &-0.33 &1   \\
Zn\,{\sc i} & 4.60 & -0.13 & 2 & & -0.09 & 3 & & -0.01 & 4 & & -0.03 
& 4 & & -0.12 & 2 & & -0.23 & 2 & & -0.43 &2 \\
Y\,{\sc ii}  & 2.21 & -0.66 & 5 & & -0.62 & 4 & & -0.08 & 7 & & -0.04 
& 4 & & -0.42 & 1 & & +0.03 & 1 & & +0.01 & 5\\
Zr\,{\sc ii} & 2.59 & \nodata & \nodata & & -0.52 & 2 & & +0.10 & 10 
& & +0.03 & 3 & & \nodata & \nodata & & \nodata & \nodata & & \nodata 
& \nodata\\
Ba\,{\sc ii} & 2.17 & -0.29 & 1 & & -0.42 & 1 & & -0.02 & 3 & & 
\nodata & \nodata & & -0.81 & 3 & & \nodata & \nodata & & +0.70 & 1 \\
La\,{\sc ii} & 1.13 & -0.59 & 1 & & -0.48 & 3 & & +0.10 & 7 & & +0.05 
& 6 & & \nodata & \nodata & & -0.41 & 2 & & +0.21 & 3\\
Ce\,{\sc ii}  & 1.58 & -0.53 & 3 & & -0.52 & 3 & & \nodata & \nodata 
& & -0.24 & 6  & & \nodata & \nodata & & +0.10 & 5 & & -0.04 & 9\\
Nd\,{\sc ii}  & 1.45 & -0.74 & 1 & & -0.60 & 1 & & \nodata & \nodata 
& & +0.05 & 14 & & \nodata & \nodata & & -0.21 & 1 & & +0.30 &6\\
Sm\,{\sc ii}  & 1.01 & \nodata & \nodata & & -0.59 & 3 & & \nodata & 
\nodata & & -0.21 & 4 & & \nodata & \nodata & & -0.22 & 1 & & +0.16 & 
2\\
Eu\,{\sc ii}  & 0.52 & -0.01 & 2 & & -0.16 & 2 & & +0.01 & 1 & & 
-0.11 & 2 & & \nodata & \nodata & & +0.15 & 1 & & +0.15 & 1\\
\hline
Fe\,{\sc i} & 7.45 & -0.17 & 99 & & -0.40 & 76 & & -0.13 & 122 & & 
+0.24 & 115 & & -1.79 & 79 & & -0.19 & 37 & & -0.45 & 91\\
Fe\,{\sc ii}& 7.45 & -0.26 & 24 & & -0.48 & 12 & & -0.15 & 38 & & 
+0.21 & 24 & & -1.86 & 17 & & -0.21 & 11 & & -0.45 & 15\\
\hline
\hline
\end{tabular}
$^a$ Solar abundances from Asplund, Grevesse, \& Sauval (2005).\\
$^b$ For each star, we give the mean value of [X/Fe] for element X 
but for Fe we give the mean [Fe/H] for the 
Fe\,{\sc i} and Fe\,{\sc ii} lines. The quantity n is in all cases the number of considered lines.

\end{table}


\setcounter{table}{2}
\begin{table}
\caption{ELEMENTAL ABUNDANCES - THE INTERMEDIATE-PERIOD 
VARIABLES}\vspace{.25cm}
\scriptsize
\begin{tabular}{lccrrrrrr@{\hspace{-.25mm}}r@{\hspace{-1mm}}r@{\hspace{-.005mm}}rrrrrrr@{\hspace{-.005mm}}rrr}

\hline
\hline
&&&&&&&&&&&&&&&&&&\\*[-.30cm]
& &  TX Del$^{b}$&& &  IX Cas &&& & &AL Vir &&&  && AP Her  \\
\cline{3-4}\cline{6-7}\cline{9-14}  \cline{16-17}

&&&&&&&&&&&&&&&&&&\\*[-.30cm]
Species & $\log \epsilon_{\odot}^{a}$&  [X/Fe] &  n &  &    [X/Fe] &
n  &   &  &  [X/Fe]$^{c}$ &  n  &  &    [X/Fe]$^{d}$ &  n  &  &
[X/Fe] &  n  &  &     \\
\hline
&&&&&&&&&&&&&&&&&&\\*[-.30cm]
C\,{\sc i}  &8.39 & -0.04 & 19 & & -1.76 & 2 & & & +0.05 & 11 & &
+0.06 & 14 & & -0.09 & 10  \\
N\,{\sc i} &  7.78 & +1.23 & 6 & &  +0.95 & 7 & & & +0.88 & 3 & &
+0.77 & 6 & & +0.92 & 6 \\
O\,{\sc i} &   8.66 & +0.36 & 3 & & +0.27 & 4 & & & +0.81 & 2 & &
+0.61 & 3 & & +0.69 & 3\\
Na\,{\sc i} & 6.17 & +0.61 & 2 & & +1.02 & 3 & & & +0.26 & 3 & &
+0.22 & 3 & & +0.29 & 1 \\
Mg\,{\sc i}  & 7.53 & +0.08 & 2 & & +0.10 & 1 & & & +0.30 & 1 & &
\nodata & \nodata & & +0.36 & 1 \\
Al\,{\sc i} &  6.37 & +0.28 & 3 & & +0.28 & 4 & & & +0.06 & 3 & &
\nodata & \nodata & & \nodata & \nodata \\
Si\,{\sc i} &  7.51 &  +0.32 & 14 & & +0.36 & 16 & & & +0.41 & 17 & &
+0.44 & 13 & & +0.53 & 11 \\
Si\,{\sc ii} & 7.51 & \nodata & \nodata & & +0.16 & 1 & & & +0.43 & 2
& & \nodata & \nodata & & +0.64 & 2 \\
S\,{\sc i}  & 7.14 & +0.27 & 6 & & +0.32 & 5 & & & +0.29 & 6 & &
+0.36 & 5 & & +0.39 & 4 \\
Ca\,{\sc i} &  6.31 & +0.07 & 5 & & +0.17 & 13 & & & +0.01 & 14 & &
+0.03 & 17 & & +0.23 & 21 \\
Sc\,{\sc ii} &  3.05 & -0.23 & 3 & & +0.11 & 8 & & & -0.37 & 8 & &
-0.32 & 9 & & -0.38 & 8\\
Ti\,{\sc i}  &  4.90 & \nodata & \nodata & & +0.17 & 2 & & & +0.15 &
4 & & \nodata & \nodata & & 0.11 & 1 \\
Ti\,{\sc ii} &  4.90 & -0.18 & 1 & & +0.17 & 11 & & & +0.10 & 8 & &
-0.02 & 18 & & 0.11 & 31\\
Cr\,{\sc i}  & 5.64 & 0.00 & 4 & & -0.07 & 11 & & & -0.17 & 13 & &
-0.23 & 11 & & -0.16 & 10 \\
Cr\,{\sc ii} & 5.64  & -0.08 & 3 & & -0.09 & 16 & & & -0.20 & 12 & &
-0.20 & 15 & & -0.16 & 16 \\
Mn\,{\sc i} & 5.39 & -0.06 & 4 & & -0.13 & 8 & & & -0.26 & 6 & &
\nodata & \nodata & & -0.21 & 5\\
Ni\,{\sc i} &  6.23 & +0.03 & 27 & & +0.04 & 35 & & & +0.07 & 43 & &
+0.06 & 36 & & +0.07 & 25 \\
Cu\,{\sc i} & 4.21  & \nodata & \nodata & & +0.13 & 1 & & & +0.14 & 1
& & \nodata & \nodata & & +0.03 & 1   \\
Zn\,{\sc i} &  4.60 &  \nodata & \nodata & &  0.00 & 4 & & & +0.14 &
3 & & 0.00 & 2 & & +0.06 & 1\\
Y\,{\sc ii}  & 2.21 & -0.29 & 4 & & -0.07 & 11 & & & -0.76 & 6 & &
-0.56 & 1 & & \nodata & \nodata \\
Zr\,{\sc ii} &  2.59 & -0.55 & 2 & & +0.03 & 11 & & & -0.54 & 2 & &
-0.38 & 3 & & -0.25 & 7 \\
Ba\,{\sc ii} & 2.17 & \nodata & \nodata & & +0.13 & 1 & & & \nodata &
\nodata & &   -0.58 &  1 & &  \nodata & \nodata\\
La\,{\sc ii} & 1.13 & -0.42 & 15 & & +0.17 & 7 & & & -0.49 & 6 & &
\nodata & \nodata & & 0.00 & 1\\
Ce\,{\sc ii}  & 1.58 & -0.83 & 3 & & -0.16 & 11 & & & -0.79 & 3 & &
-0.70 & 1 & & \nodata & \nodata \\
Nd\,{\sc ii}  & 1.45 & -0.49 & 9 & & +0.02 & 9 & & & -0.49 & 7 & & \nodata
& \nodata & & \nodata & \nodata  \\
Sm\,{\sc ii}  &1.01 & \nodata & \nodata & & -0.19 & 4 & & & -0.35 & 5
&  & \nodata  & \nodata & & \nodata & \nodata \\
Eu\,{\sc ii}  & 0.52 & -0.39 & 1 & & +0.27 & 2 & & & +0.11 & 2 & & 
\nodata & \nodata
& & +0.39 & 2 \\
\hline
Fe\,{\sc i} & 7.45 & +0.11 & 85 & & -0.48 & 108 & & & -0.38 & 85 & &
-0.41 & 92 & & -0.75 & 103 \\
Fe\,{\sc ii}& 7.45 & +0.11 & 27 & & -0.53 & 34 & & & -0.41 & 27 & &
-0.44 & 24 & & -0.80 & 36 \\
\hline
\hline
\end{tabular}

$^a$ Solar abundances from Asplund, Grevesse, \& Sauval (2005).\\
$^b$ For each star, we give the mean value of [X/Fe] for element X
but for Fe we give the mean [Fe/H] for the
Fe\,{\sc i} and Fe\,{\sc ii} lines. \\ The quantity n is in all cases the
number of considered lines.\\
$^c$ Analysis for 2004 May   3\\
$^d$ Analysis for 2005 January 30

\end{table}

\setcounter{table}{3}
  \begin{table}
  \caption{ELEMENTAL ABUNDANCES - THE W VIR VARIABLES}\vspace{.25cm}
\tiny
  \begin{tabular}{lcc@{\hspace{-2mm}}rr@{\hspace{-.25mm}}r@{\hspace{-.25mm}}rr@{\hspace{-.25mm}}r@{\hspace{-.25mm}}rr@{\hspace{-.25mm}}r@{\hspace{-.25mm}}rr@{\hspace{-.25mm}}r@{\hspace{-5mm}}rr@{\hspace{-2mm}}rrr@{\hspace{-.25mm}}r@{\hspace{-.25mm}}r@{\hspace{-1mm}}rr@{\hspace{-3mm}}r@{\hspace{-2mm}}rr@{\hspace{1mm}}r@{\hspace{-3mm}}rrr}

  \hline
  \hline
  &&&&&&&&&&&&&&&&&&&&&&\\*[-.20cm]
  & &  CO Pup$^{b}$&& & SZ Mon  &&& W Vir &&& MZ Cyg  &&& &  CC Lyr  &
  & &&  & RX Lib &&&  & TW Cap & & & & & V1711 Sgr\\
  \cline{3-4}\cline{6-7}\cline{9-10} \cline{12-13} \cline{15-19}
  \cline{21-22}\cline{24-28} \cline{30-31}

  &&&&&&&&&&&&&&&&&&&&&&\\*[-.20cm]
  Species & $\log \epsilon_{\odot}^{a}$&  [X/Fe] &  n &  &    [X/Fe] &
  n  &   &   [X/Fe] &  n  &  &    [X/Fe] &  n  &  &     [X/Fe]$^{c}$ &
  n  &  &    [X/Fe]$^{d}$ &  n  &  &    [X/Fe] &  n  &  &    [X/Fe]$^{e}$ &
  n  &  &    [X/Fe]$^{f}$ &  n & &  [X/Fe] &  n\\
  \hline
  &&&&&&&&&&&&&&&&&&&&&&\\*[-.20cm]
  C\,{\sc i} & 8.39 & +0.23 & 3  & & +0.57 & 3 & & -0.23 & 4 & & +0.73
  & 4 & & +3.37 & 4 & & +3.58 & 18 & & +0.19 & 1  & &
  \nodata & \nodata & & $<$-0.1 & 1 & & +0.81 & 5\\
  N\,{\sc i} & 7.78 & +1.33 & 2 & & \nodata & \nodata & & \nodata &
  \nodata & &  +0.39: &2  & & +4.04 & 1 & & \nodata & \nodata && +1.12 
& 2  & & \nodata & \nodata & & $<$+0.8 & 2 & &
  +0.42 & 1\\
  O\,{\sc i} & 8.66 & +0.29 & 2 & & +0.04 & 2 & & +0.74 & 3 & & +0.18
  & 2 & & +3.96 & 1 & & +3.96 & 3 & & +0.49 & 3  & & +0.70
  & 2 & &  $<$+0.5 & 1 & & +0.95 & 3\\
  Na\,{\sc i} & 6.17 & +0.15 & 4 & & +0.44 & 2 & & +0.36 & 3 & & +0.09
  & 3 & & +2.75 & 1 & & +3.04 & 1 & & +0.02 & 1  & &
  \nodata & \nodata & & \nodata&\nodata  & & +0.60 & 4\\
  Mg\,{\sc i}  & 7.53 & +0.18 & 4 & & +0.33 & 3 & & +0.67 & 5 & & -0.02
  & 3 & & +0.34 & 3 & & +0.54 & 3 & & +0.85 & 3  & &+0.45
  & 2 & & +.38 & 2 & & +0.59 & 4\\
  Al\,{\sc i} & 6.37 & -0.45 & 3 & & -0.85 & 1 & & -0.13 & 3 & & -0.58
  & 3 & & \nodata &   & & \nodata &\nodata & & -0.15 & 1  & & \nodata 
& \nodata &
  & -0.87 & 1& & -0.95 & 1 \\
  Si\,{\sc i} & 7.51 &  +0.19 & 5 & & +0.37 & 8 & & +0.55 & 8 & & +0.27
  & 7 & &\nodata & && \nodata&\nodata && +0.53 & 5  &  & +0.64 & 6 & & +0.52 &
  2 & & +0.46 & 8\\
  Si\,{\sc ii} & 7.51 & +0.30 & 2 & & \nodata & \nodata & & +0.71 & 2 &
  & \nodata & \nodata & &\nodata &  & &\nodata & \nodata && +0.57 & 2 & &
  +0.46 & 2 & & +0.48 & 2  & & +0.87 & 2 \\
  S\,{\sc i}  & 7.14 & +0.55 & 2 & & +0.64 & 2 & & +0.74 & 4 & & +0.73
  & 4 & & +3.20 & 4 & & +3.60 & 5 & & +0.42 & 4  & & +0.40
  & 1 & & +0.33 & 1& & +1.13 & 4\\
  Ca\,{\sc i} & 6.31 & -0.37 & 6 & & +0.21 & 3 & & -0.03 & 8 & & -0.45
  & 6 & &\nodata &  & & \nodata&\nodata & & -0.05 & 5  & &+0.35 & 18 & & +0.29
  & 10 & &  -0.12 & 9 \\
  Sc\,{\sc ii} & 3.05 & -0.96 & 3 & & -1.06 & 2 & & -1.03 & 4 & & -0.87
  & 5 & & \nodata & & &\nodata &\nodata & & -1.22 & 5  & & +0.19 & 6 & & -0.14 &
  8 & & -1.26 & 3 \\
  Ti\,{\sc i}  & 4.90 &  -0.39 & 8 & & \nodata & \nodata & & -0.17 & 9
  & & -0.51 & 3 & &\nodata & & &\nodata & \nodata & &-0.04 & 4  & &
  +0.38 & 3 & & \nodata & \nodata & & \nodata &  \\
  Ti\,{\sc ii} & 4.90 & -0.28 & 9 & & -0.85 & 5 & & +0.02 & 6 & & -0.35
  & 3 & &\nodata & & & \nodata &\nodata & & -0.24 & 4 & & +0.37 & 24 & & +0.04
  & 5  & & -0.85 & 4 \\
  Cr\,{\sc i}  & 5.64 & -0.49 & 9 & & -0.04 & 3 & & -0.47 & 9 & & -0.03
  & 2 & &\nodata & & & \nodata&\nodata & & -0.33 & 5 & & -0.18 & 6 & & +0.01 &
  2 & & -0.40 & 9 \\
  Cr\,{\sc ii} & 5.64 & -0.34 & 5 & & -0.13 & 7 & & -0.26 & 4 & & -0.11
  & 4 & &\nodata & & & \nodata& \nodata& & -0.36 & 4  & & -0.13 & 9 & & -0.15
  & 4 & & -0.13 & 17 \\
  Mn\,{\sc i} & 5.39 & -0.35 & 6 & & -0.18 & 4 & & -0.46 & 5 & & -0.39
  & 3 & & +1.00 & 2 & &\nodata &\nodata & & -0.30 & 2  & & -0.45 & 3 & &
  \nodata & \nodata & & -0.16 & 5\\
  Ni\,{\sc i} & 6.23 &  -0.12 & 24 & & -0.11 & 7 & & -0.05 & 24 & &
  -0.15 & 11 & &\nodata & & &\nodata & \nodata& & +0.09 & 10  & & +0.06 & 16 &
  & -0.14 & 2 & & -0.12 & 7 \\
  Cu\,{\sc i} & 4.21 & -0.27 & 2 & & -0.15 & 1 & & -0.25 & 2 & &
-0.01 & 1  & &\nodata & & &\nodata & \nodata& &  &   & & -0.28 & 1  &
  &  & & & -0.03 & 2 \\
  Zn\,{\sc i} & 4.60 & -0.13 & 3 & & -0.03 & 2 & & +0.16 & 4 & & -0.03
  & 3 & & +3.05 & 2 & & +3.01 & 2 & & +0.20 & 4  & & +0.28
  & 2 & & +0.01 & 2 & & +0.22 & 4\\
  Y\,{\sc ii}  & 2.21 & -1.53 & 1 & & -1.42 & 1 & & -1.92 & 1 & & -0.80
  & 2 & &\nodata & & &\nodata & \nodata& & -0.38 & 1 & & -0.08 & 3 & & -0.11
  & 1& &  -0.26 & 1 \\
  Zr\,{\sc ii} & 2.59 & \nodata & \nodata & & \nodata & \nodata & &
  \nodata & \nodata & & \nodata & \nodata & & \nodata& & & \nodata& 
\nodata& & \nodata &
  \nodata  & & -0.01 & 2 & & -0.24 & 1  & & \nodata &  \\
  Ba\,{\sc ii} & 2.17 &  -0.40 & 1 & & \nodata & \nodata & & -0.36 & 2
  & & -0.90   & 1 & &\nodata & & & \nodata& \nodata& &  -0.42 & 2 & &
  \nodata & \nodata & & -0.10 & 2 & & -0.10 & 2  \\
  La\,{\sc ii} & 1.13 &  -0.83 & 2 & & -0.96 & 3 & & -0.54 & 3 & &
  -1.32 & 1 & &\nodata & & &\nodata &\nodata & & -0.09 & 1  & & +0.22 &
  4 & & \nodata & \nodata & & \nodata &  \\
  Ce\,{\sc ii}  & 1.58 & -0.72 & 5 & & \nodata & \nodata & & -0.69 & 7
  & & -0.97 & 3 & & \nodata& & & \nodata& \nodata& & -0.70 & 1  & & 
+0.08 & 4 & & \nodata & \nodata & & -0.23 & 7\\
  Nd\,{\sc ii}  & 1.45 & -0.58 & 4 & & \nodata & \nodata & & -0.53 & 5
  & & -0.83   & 4       & & \nodata& & & \nodata& \nodata& &\nodata 
&\nodata  & & \nodata & \nodata  &
  & \nodata & \nodata & & -0.39 & 1\\
  Sm\,{\sc ii}  & 1.01 & -0.40 & 2 & & -0.31 & 1 & & -0.55 & 4 & &
  -1.03   & 1 & &\nodata & & &\nodata &\nodata & & \nodata&\nodata& & \nodata &
  \nodata & & \nodata & \nodata  & & -0.28 & 1 \\
  Eu\,{\sc ii}  & 0.52 &  -0.11 & 2 & & -0.37 & 1 & & +0.12 & 1 & &
  -0.27 & 1 & &\nodata & & &\nodata &\nodata & &  -0.01 & 2 & & +0.05 
& 2 & & \nodata & \nodata& & +0.34 & 1 \\
  \hline
  Fe\,{\sc i} & 7.45 &  -0.64 & 130 & & -0.43 & 59 & & -0.97 & 110 & &
  -0.18 & 56 & & -3.62 & 2 & & -4.11 & 2 & & -0.95 & 61
  & & -1.75 & 91 & & -1.79 & 33& & -1.17 & 67 \\
  Fe\,{\sc ii}& 7.45 &  -0.57 & 12 & & -0.50 & 8 & & -0.93 & 22 & &
  -0.24 & 9 & & -3.72 & 1 & & \nodata & \nodata & & -1.02 & 14  & & 
-1.72 & 31 & & -1.81 & 13 & &
  -1.17 & 15\\
  \hline
  \hline
  \end{tabular}
  $^a$ Solar abundances from
  Asplund, Grevesse, \& Sauval (2005).\\
$^b$ For each
   star, we give the mean value of [X/Fe] for element X
but for Fe we give the mean [Fe/H] for the
Fe\,{\sc i} and Fe\,{\sc ii} lines. The quantity n is in all cases the
number of considered lines.
  $^c$ Analysis for 2004 August 9\\
  $^d$ Analysis for 2005 March 7\\
  $^e$ Analysis for 2004 December 8\\
  $^f$  Analysis for 2005 July 3

  \end{table}

\end{document}